\documentclass[sigplan,screen, 10pt]{acmart}
\usepackage{subcaption}
\usepackage{balance}
\usepackage{listings}
\usepackage{booktabs}
\usepackage{color}
\newif\ifshowstructure
\newcommand{\structurebeginitemize}{\ifshowstructure \begin{itemize} \fi}
\newcommand{\structureenditemize}{\ifshowstructure \end{itemize} \fi}
\newcommand{\structureitem}{\ifshowstructure \item \fi}
\newcommand{\structureitemwithtext}[1] {\ifshowstructure \item #1 \fi}

\makeatletter
\newcommand\footnoteref[1]{\protected@xdef\@thefnmark{\ref{#1}}\@footnotemark}
\makeatother

\showstructurefalse

\usepackage{tikz}
\usepackage{amsmath}
\DeclareMathOperator*{\argmin}{arg\,min}
\usepackage{multirow}
\usepackage{comment}

\usepackage{color}
\definecolor{lightgray}{rgb}{.9,.9,.9}
\definecolor{darkgray}{rgb}{.4,.4,.4}
\definecolor{purple}{rgb}{0.65, 0.12, 0.82}
\definecolor{xyellow}{rgb}{1, 0.89, 0.71}
\definecolor{applegreen}{rgb}{0.0, 0.5, 0.0}
\lstdefinelanguage{JavaScript}{
  keywords={break, case, catch, continue, debugger, default, delete, do, else, false, finally, for, function, if, in, instanceof, new, null, return, switch, this, throw, true, try, typeof, var, void, while, with},
  morecomment=[l]{//},
  morecomment=[s]{/*}{*/},
  morestring=[b]',
  morestring=[b]",
  ndkeywords={require, cpuUsage, handler, putItem},
  keywordstyle=\color{blue}\bfseries,
  ndkeywordstyle=\color{applegreen}\bfseries,
  identifierstyle=\color{black},
  commentstyle=\color{purple}\ttfamily,
  stringstyle=\color{xyellow}\ttfamily,
  sensitive=true
}

\begin{document}

\title{Sizeless: Predicting the Optimal Size of Serverless Functions}

\author{Simon Eismann}\affiliation{\institution{University of W\"urzburg}\country{Germany}}
\author{Long Bui}\affiliation{\institution{University of W\"urzburg}\country{Germany}}
\author{Johannes Grohmann}\affiliation{\institution{University of W\"urzburg}\country{Germany}}
\author{Cristina L. Abad}\affiliation{\institution{ESPOL}\country{Ecuador}}
\author{Nikolas Herbst}\affiliation{\institution{University of W\"urzburg}\country{Germany}}
\author{Samuel Kounev}\affiliation{\institution{University of W\"urzburg}\country{Germany}}

\begin{abstract}
Serverless functions are a cloud computing paradigm where the provider takes care of resource management tasks such as resource provisioning, deployment, and auto-scaling. The only resource management task that developers are still in charge of is selecting how much resources are allocated to each worker instance. However, selecting the optimal size of serverless functions is quite challenging, so developers often neglect it despite its significant cost and performance benefits. Existing approaches aiming to automate serverless functions resource sizing require dedicated performance tests, which are time-consuming to implement and maintain.

In this paper, we introduce an approach to predict the optimal resource size of a serverless function using monitoring data from a single resource size. As our approach does not require dedicated performance tests, it enables cloud providers to implement resource sizing on a platform level and automate the last resource management task associated with serverless functions. We evaluate our approach on four different serverless applications on AWS, where it predicts the execution time of the other memory sizes based on monitoring data for a single memory size with an average prediction error of 15.3\%. Based on these predictions, it selects the optimal memory size for 79.0\% of the serverless functions and the second-best memory size for 12.3\% of the serverless functions, which results in an average speedup of 39.7\% while also decreasing average costs by 2.6\%.
\end{abstract}

\maketitle

\section{Introduction}
\begin{figure*}
    \centering
    \begin{subfigure}[b]{0.47\textwidth}
        \centering
        \includegraphics[width=\textwidth]{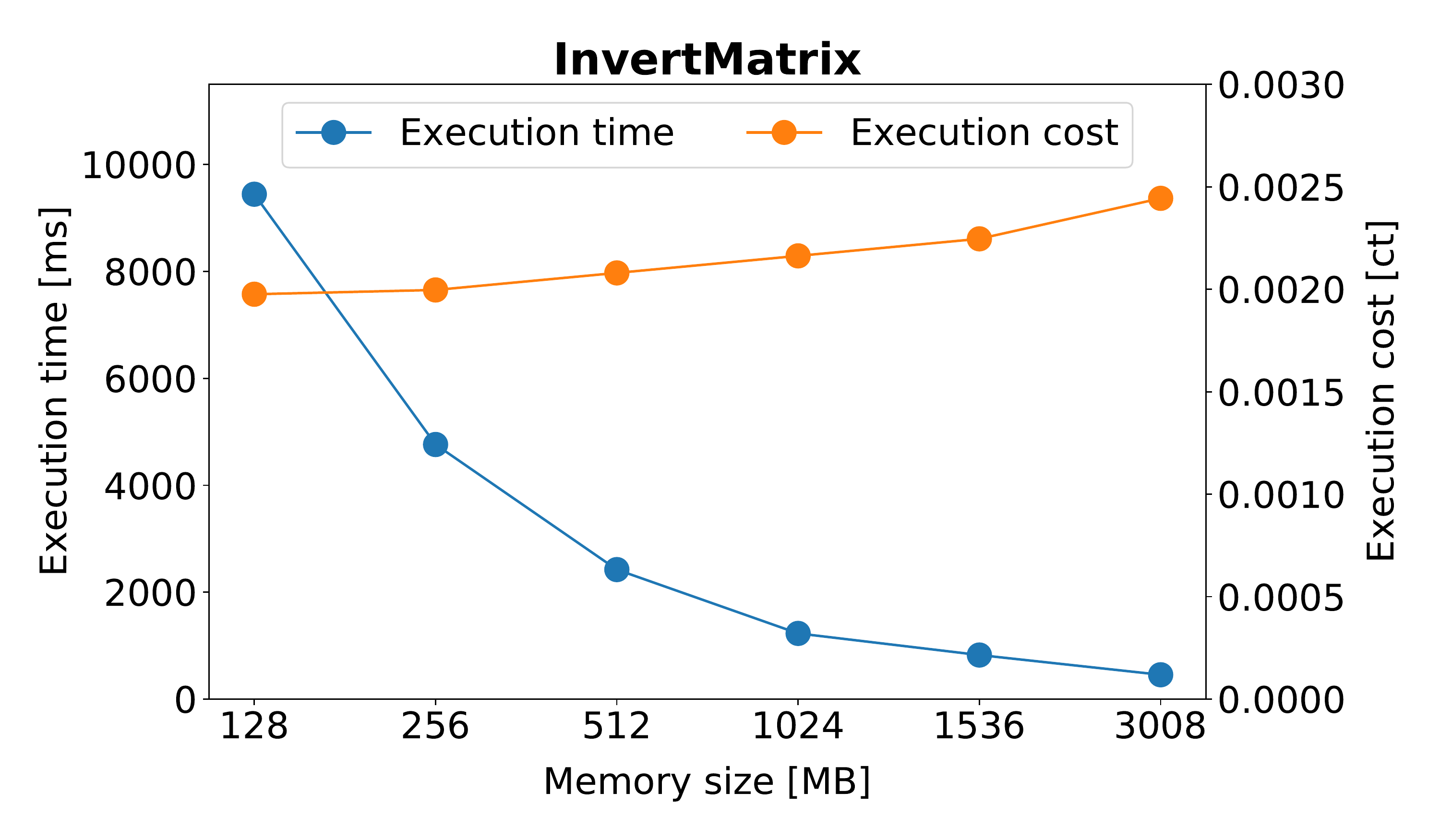}
    \end{subfigure}
    \begin{subfigure}[b]{0.47\textwidth}  
        \centering 
        \includegraphics[width=\textwidth]{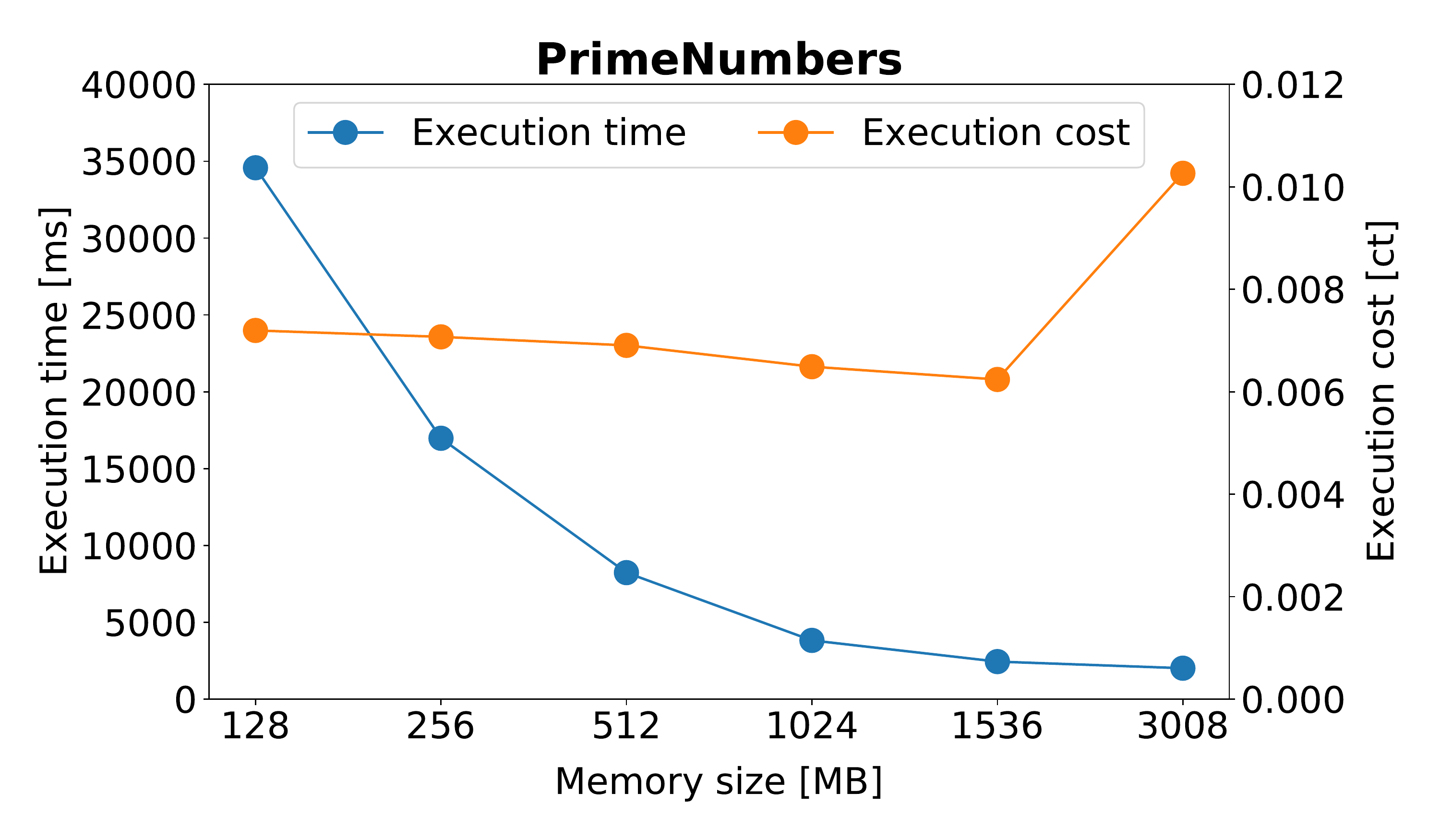}
    \end{subfigure}
    \begin{subfigure}[b]{0.47\textwidth}   
        \centering 
        \includegraphics[width=\textwidth]{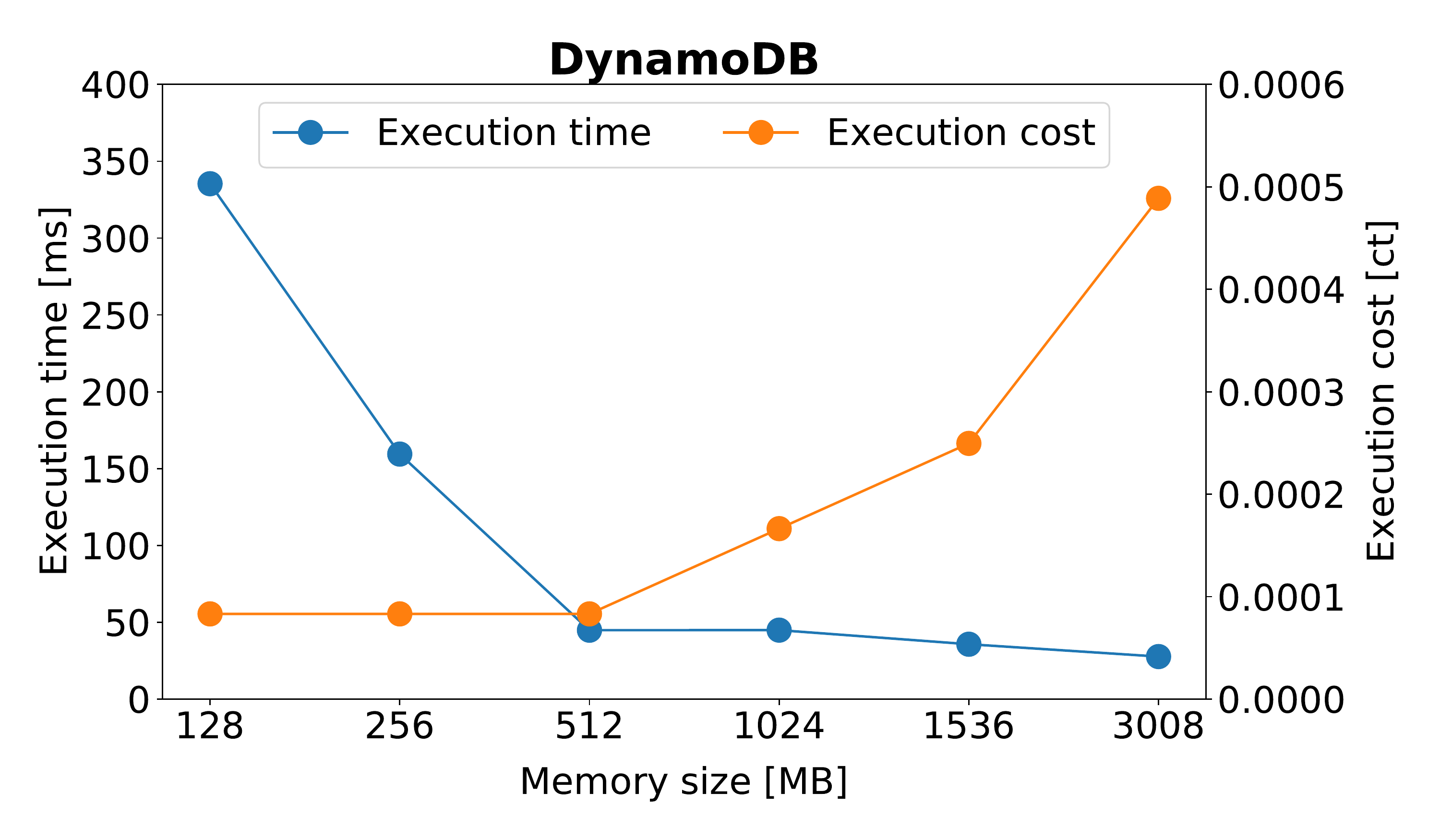}
    \end{subfigure}
    \begin{subfigure}[b]{0.47\textwidth}   
        \centering 
        \includegraphics[width=\textwidth]{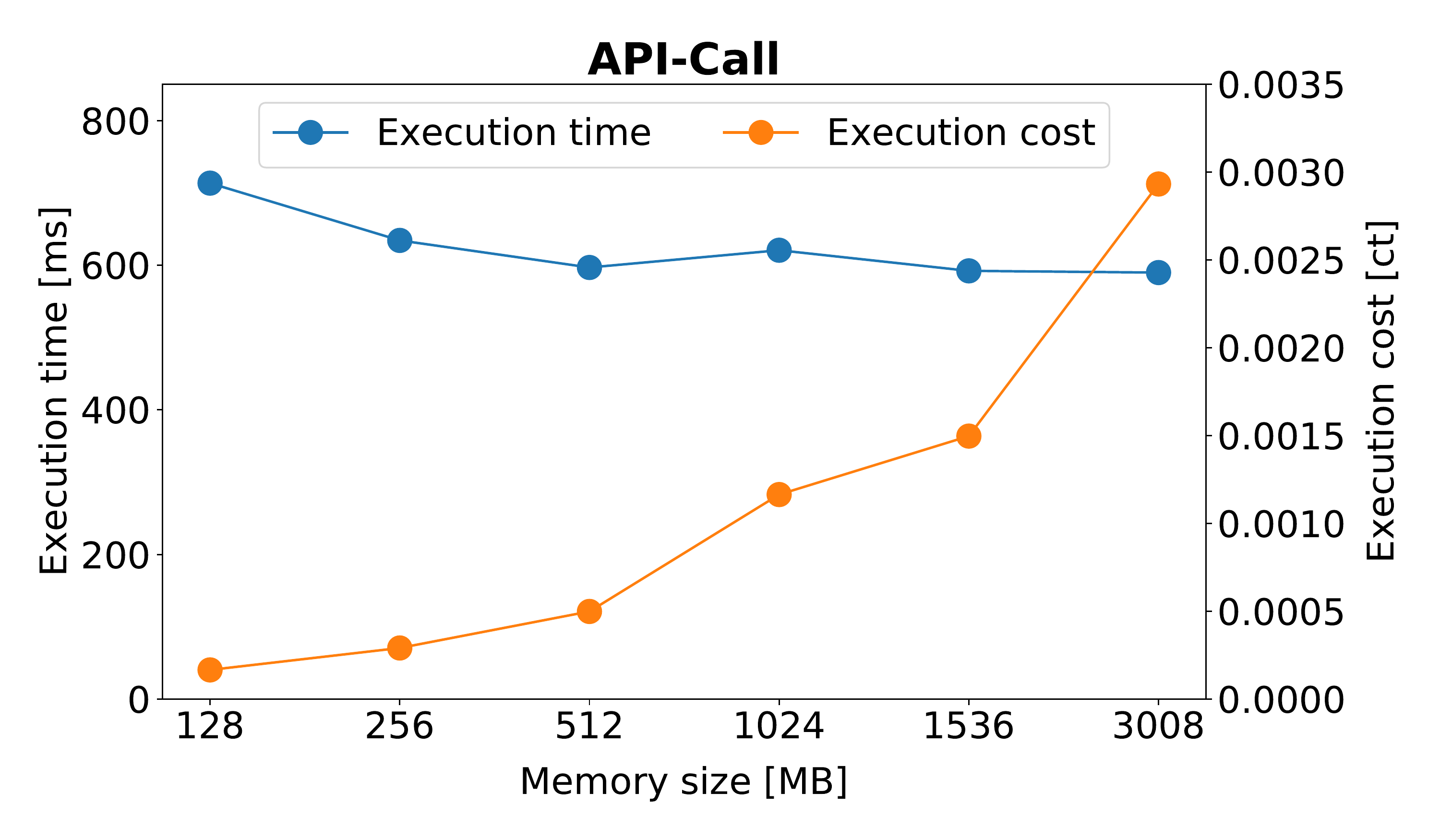}
    \end{subfigure}
    \caption{The mean execution time and cost for four serverless functions~(adapted from~\cite{Casalboni2020Deep}).} 
    \label{fig:motivatingExample}
\end{figure*}

\structurebeginitemize
    \structureitemwithtext{Context}
        \structurebeginitemize
            \structureitem Serverless computing is an emerging cloud computing paradigm that combines managed services (databases, storage, authentication, pub/sub, queueing, etc.) with so-called serverless functions~\cite{jonas2019cloud, Eyk2018Serverless}.
            \structureitem These \emph{serverless functions} are a managed compute solution that executes code in response to event triggers or HTTP requests~\cite{castro2019rise, lynn2017preliminary}.
            \structureitem The main benefit of serverless functions compared to traditional compute solutions is that the cloud provider opaquely takes care of common resource management tasks such as resource provisioning, deployment, or auto-scaling~\cite{adzic2017serverless, eyk2019Reference}.
            \structureitem  With serverless functions, developers no longer need to think about resource management and are billed based on a  pay-per-use basis~\cite{eivy2017wary}.
        \structureenditemize
        
    \structureitemwithtext{Problem}
        \structurebeginitemize
            \structureitem However, there is still one resource management task that cloud providers leave to developers: \textit{resource sizing}. 
            \structureitem Resource sizing is the task of selecting how much CPU, memory, I/O bandwidth, etc. are allocated to a worker instance~\cite{singh2020auto, Agarwal2019cost, Piraghaj2015Efficient}.
            \structureitem Most cloud providers implement the resource sizing of serverless functions as a configurable memory size, where other resources such as CPU, network, or I/O are scaled accordingly~\cite{awspricing, gcloudpricing}.
            \structureitem Selecting an appropriate resource size is essential as it can often result in a faster execution at a lower cost.
            \structureitem However, selecting an appropriate resource size is challenging. A recent survey revealed that 47\% of the serverless functions in production have the default memory size, indicating that developers often neglect resource sizing~\cite{Datadog}.
        \structureenditemize 
        
    \structureitemwithtext{Related Work}
        \structurebeginitemize
            \structureitem There are many articles describing the impact of memory size on serverless functions~\cite{serverlessVideoProc, PeekingBehindTheCurtains, figiela2018performance, back2018using, Strehl2018Benchmark, scheuner2020function}, but existing approaches for the cost optimization of serverless functions mostly do not consider memory size~\cite{costradamus, spock, predictingServerlessWorkflows, costless, serverlessSimBudget}.
            \structureitem The existing approaches to optimize the memory size of serverless functions~\cite{Casalboni2020Power, Akhtar2020COSE, ali2020batch} require automated performance tests, which are time-consuming to implement and maintain~\cite{jiang2015survey, Bezemer2019DevOpsSurvey}.
            
    \structureitemwithtext{Approach}
        \structurebeginitemize
            \structureitem In this paper, we introduce an approach to predict the optimal memory size of serverless functions based on monitoring data for a single memory size.
            \structureitem First, we implement a serverless function generator capable of generating a large number of synthetic serverless functions by combining representative function segments.
            \structureitem Next, we measure the execution time and resource consumption metrics of 2\,000 synthetic functions for six different memory sizes on a public cloud.
            \structureitem Finally, we construct a multi-target regression model to predict the execution time of a serverless function for previously unseen memory sizes based on the execution time and resource consumption metrics for a single memory size.
        \structureenditemize
        
    \structureitemwithtext{Benefit}
        \structurebeginitemize
            \structureitem Unlike existing approaches based on performance testing, our approach only requires monitoring data that can be collected in production as opposed to dedicated performance tests. 
            \structureitem For developers, this removes the effort required to implement and maintain representative performance tests.
            \structureitem For cloud providers, it enables memory size recommendations, similar to the AWS Compute Optimizer for virtual machines~\cite{computeOptimizer}, which so far was infeasible as cloud providers cannot run performance tests on user functions.
        \structureenditemize
        
    \structureitemwithtext{Evaluation}
        \structurebeginitemize
            \structureitem To evaluate if our model --- which was trained on data from synthetic functions --- can be transferred to realistic serverless functions, we apply it to four serverless applications on AWS, the last of which was measured nine months after the training dataset was collected.
            Based on monitoring data from a single memory size, our approach predicts the execution time of the other memory sizes with an average prediction error of 15.3\%. It selects the optimal memory size for 79.0\% and the second-best memory size for 12.3\% of the serverless functions. Using the memory sizes selected by our approach results in an average speedup of 39.7\% while simultaneously decreasing average costs by 2.6\%.
        \structureenditemize 
    
    \structureitemwithtext{Remainder}
        \structurebeginitemize
           \structureitem The remainder of this paper is structured as follows:
           \structureitem Section~\ref{MotivatingExample} introduces a motivating example highlighting the importance and difficulty of selecting an appropriate memory size.
          \structureitem Section~\ref{approach} describes the proposed approach and Section~\ref{Evaluation} details our evaluation of the proposed approach.
           \structureitem Section~\ref{Limitations} discusses the limitations of the approach
           \structureitem Section~\ref{RelatedWork} describes the existing work on the memory size optimization of serverless functions and Section~\ref{ReplicationPackage} contains our replication package.
           \structureitem Finally, Section~\ref{Conclusion} concludes the paper by discussing potential future work.
           \structureitem 
       \structureenditemize 
\structureenditemize

\section{Motivating Example}
\label{MotivatingExample}
The relationship between the memory size of a serverless function, the cost per function execution, and the function execution time is quite counter-intuitive. A common assumption is that a higher memory size results in a faster execution at a higher price, since the allocated CPU, I/O, network, etc. capacity scales linearly with the selected memory size~\cite{awspricing, gcloudpricing}. However, this is not the case due to the pricing scheme most cloud providers employ, where the cost of an execution is calculated based on the consumed GB-s of memory, that is, the execution time multiplied by memory size. Additionally, cloud providers often include a static overhead charge per execution, which tends to be negligible compared to the consumed GB/s charge. For a more in-depth discussion of the pricing model behind serverless functions, we refer to~\cite{eivy2017wary}. As an example, a function that runs on AWS for three seconds with a memory size of 512 MB would cost: 

\begin{figure*}
    \centering
    \includegraphics[width=0.85\textwidth]{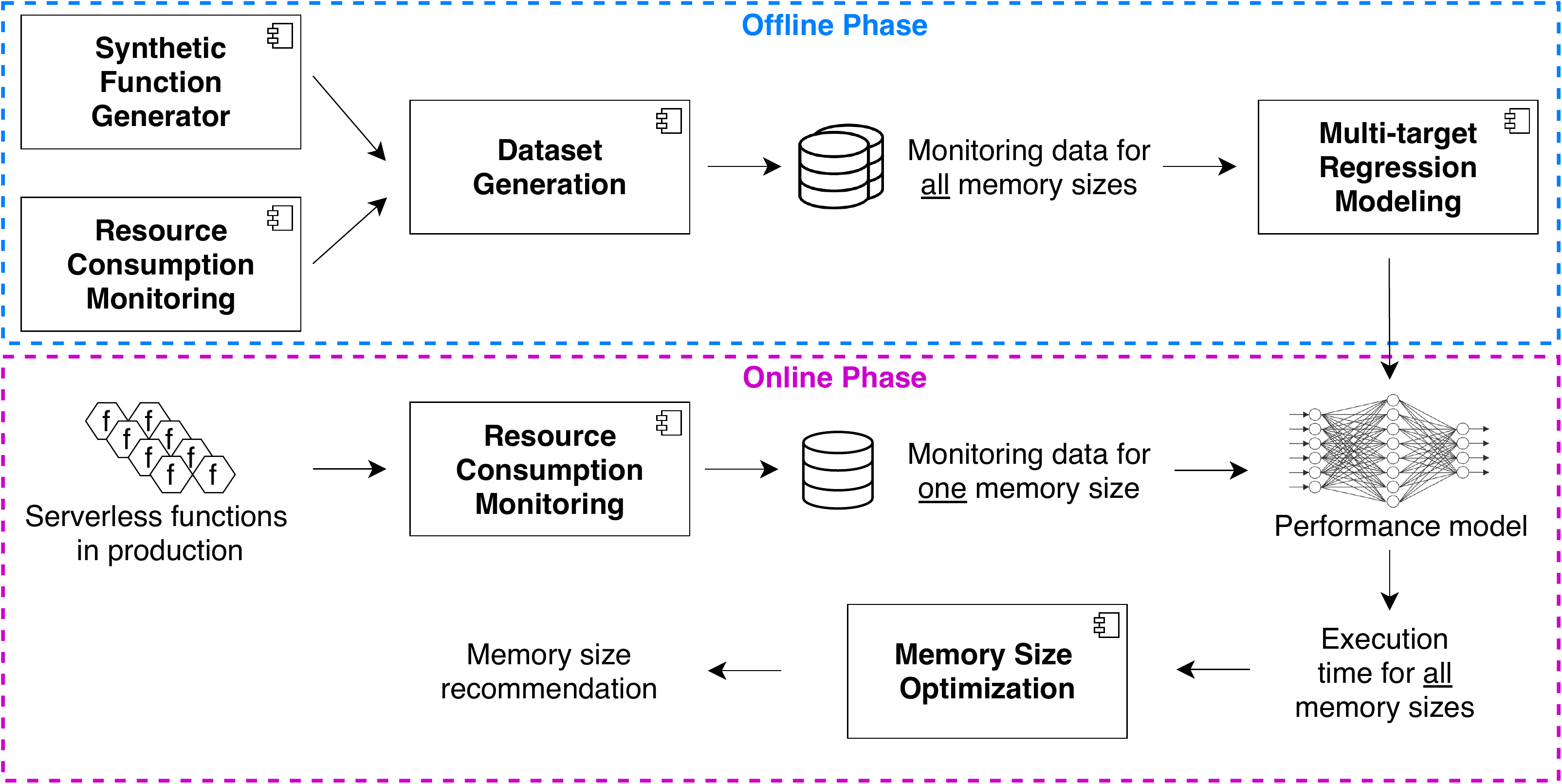}
    \caption{Overview of the proposed approach.}
    \label{fig:over}
\end{figure*}

\begin{equation*}
    3 s \cdot 0.5 GB \cdot 0.00001667\$ + 0.0000002\$ = 0.0000252\$
\end{equation*}
where 0.00001667\$ is the AWS-specific price per consumed GB-s and 0.0000002\$ the static overhead charge (0.7\% of the total execution cost). Increasing the memory size increases the cost per second, but also decreases the execution time more resources are allocated. As each function's execution time scales differently with additional resources, every function has a unique cost/performance trade-off. 

Figure~\ref{fig:motivatingExample} exemplifies how the execution time and cost per execution vary for four different functions with different memory sizes based on data from~\cite{Casalboni2020Deep}. The function \textit{InvertMatrix} creates and inverts a random matrix. Here, we can see that increasing the memory size from 128MB to 256MB decreases the execution time by 49.6\%, with only a 1\% increase in cost. For larger memory sizes, the execution time still decreases almost linearly. 
The second function, \textit{PrimeNumbers}, calculates the first million prime numbers a thousand times, which is another CPU-intensive task. Interestingly, the execution time of this function scales super-linearly with increased memory sizes up to 2048MB, which results in a 92.9\% faster execution with simultaneously 13.3\% reduced costs. Using a memory size of 3008MB further speeds up the execution time, but it increases the execution cost.
The third function, \textit{DynamoDB}, executes three queries against a DynamoDB table, which is a serverless database. Here, the execution time decreases roughly linearly from 128MB to 512MB, resulting in an 86.6\% decreased execution time at a similar cost. However, further increasing the memory only slightly reduces the execution time while increasing costs by 587.5\%.
Lastly, the \textit{API-Call} function calls an external API. Here, increasing the memory has minimal impact on the execution time and only increases the cost per execution.

Based on these results, we can conclude that: i) the impact of memory size configurations on execution time differs from function to function, ii) predicting the execution time for a memory size is challenging, as even two seemingly CPU-intensive and two network-intensive functions behave differently, and iii) selecting an appropriate memory size is important as it can drastically improve performance at a similar or reduced cost.

\section{Approach}
\label{approach}
Figure~\ref{fig:over} gives a graphical overview of our approach to predict the optimal memory size of serverless functions based on monitoring data collected for a single memory size. During the offline phase, the \textit{Synthetic Function Generator} creates many synthetic serverless functions, which are then instrumented with our \textit{Resource Consumption Monitoring}. During the \textit{Dataset Generation}, we run performance tests to obtain the resource consumption metrics and execution times for all memory sizes of thousands of synthetic serverless functions. By applying \textit{Multi-target Regression Modeling} to the resulting dataset, we generate a performance model that can predict the execution time for all memory sizes of a real function based on monitoring data for a single memory size. The \textit{Memory Size Optimization} utilizes these predictions to determine the optimal memory size. Our implementation of the proposed approach is limited to AWS Lambda and the language Node.js as they are by far the most common platform and programming language for serverless functions~\cite{eismann2020serverless, Daly2020Community}. However, we are confident that it can be transferred to other platforms and programming languages. 

\begin{table}[t]
\caption{Metric sources and collected metrics}
\begin{tabular}{@{}lll@{}}
\toprule
Metric Name            & Metric Source              \\ \midrule
Execution time         & process.hrtime()           \\
User CPU time          & process.cpuUsage()         \\
System CPU time        & process.cpuUsage()         \\
Vol Context Switches   & process.resourceUsage()    \\
Invol Context Switches & process.resourceUsage()    \\
File system reads      & process.resourceUsage()    \\
File system writes     & process.resourceUsage()    \\
Resident set size      & process.memoryUsage()      \\
Max resident set size  & process.resourceUsage()    \\
Total heap             & process.memoryUsage()      \\
Heap used              & process.memoryUsage()      \\
Physical heap          & v8.getHeapStatistics()     \\
Available heap         & v8.getHeapStatistics()     \\
Heap limit             & v8.getHeapStatistics()     \\
Allocated memory       & v8.getHeapStatistics()     \\
External memory        & process.memoryUsage()      \\
Bytecode metadata      & v8.getHeapCodeStatistics() \\
Bytes received         & /proc/net/dev/             \\
Bytes transmitted      & /proc/net/dev/             \\
Packages received      & /proc/net/dev/             \\
Packages  transmitted  & /proc/net/dev/             \\
Min event loop lag     & perf\_hooks    \\
Max event loop lag     & perf\_hooks    \\
Mean event loop lag    & perf\_hooks    \\
Std event loop lag     & perf\_hooks    \\ \bottomrule
\end{tabular}
\label{tbl:metrics}
\end{table}

\subsection{Synthetic function generator}
\label{a1}
Learning to derive how different memory sizes influence the execution time based on resource consumption metrics requires a large dataset covering many different functions. Unfortunately, there are not enough easily benchmarkable open-source functions available~\cite{eismann2020review}. Therefore, we propose to generate synthetic serverless functions by combining representative function segments.
Each function segment represents the smallest granularity of common tasks in serverless functions. Additionally, each function segment has to provide its own inputs to simplify load generation (e.g., a function segment that performs image manipulation comes with several images). Further, each function segment provides setup and teardown code of all external services it uses (e.g., databases or messaging queues).

For the selection of function segments, we investigated common tasks from a recent survey of 89 serverless applications~\cite{eismann2020serverless}. We implemented sixteen function segments covering, among other things, CPU-intensive tasks, image manipulation, format conversion, data compression, and interaction with files and external services, such as DynamoDB or S3. The function generator documentation from our replication package includes detailed descriptions of all implemented function segments\footnote{\label{r1}\url{https://github.com/Sizeless/ReplicationPackage}}.
While it is not possible to cover all possible functionality serverless functions can implement, our function segments should enable the generation of a large number of synthetic serverless functions with varying resource consumption profiles. In the future, the number of implemented segments can be easily extended if specific resource consumption profiles are missing.

The function generator randomly combines these function segments and wraps them in a Lambda handler. The resulting deployment package and \textit{template.yaml} file can be deployed as an AWS CloudFormation stack using the Serverless Application Model. Additionally, it generates scripts to setup and teardown the required services from the segments. The function generator keeps a list of already generated function hashes to ensure that no function is generated twice. For additional details on the function generator implementation, we refer to the documentation in our replication package\footnoteref{r1}. 

\begin{figure}[t]
    \centering
    \includegraphics[width=\linewidth, trim=0 24 0 0, clip]{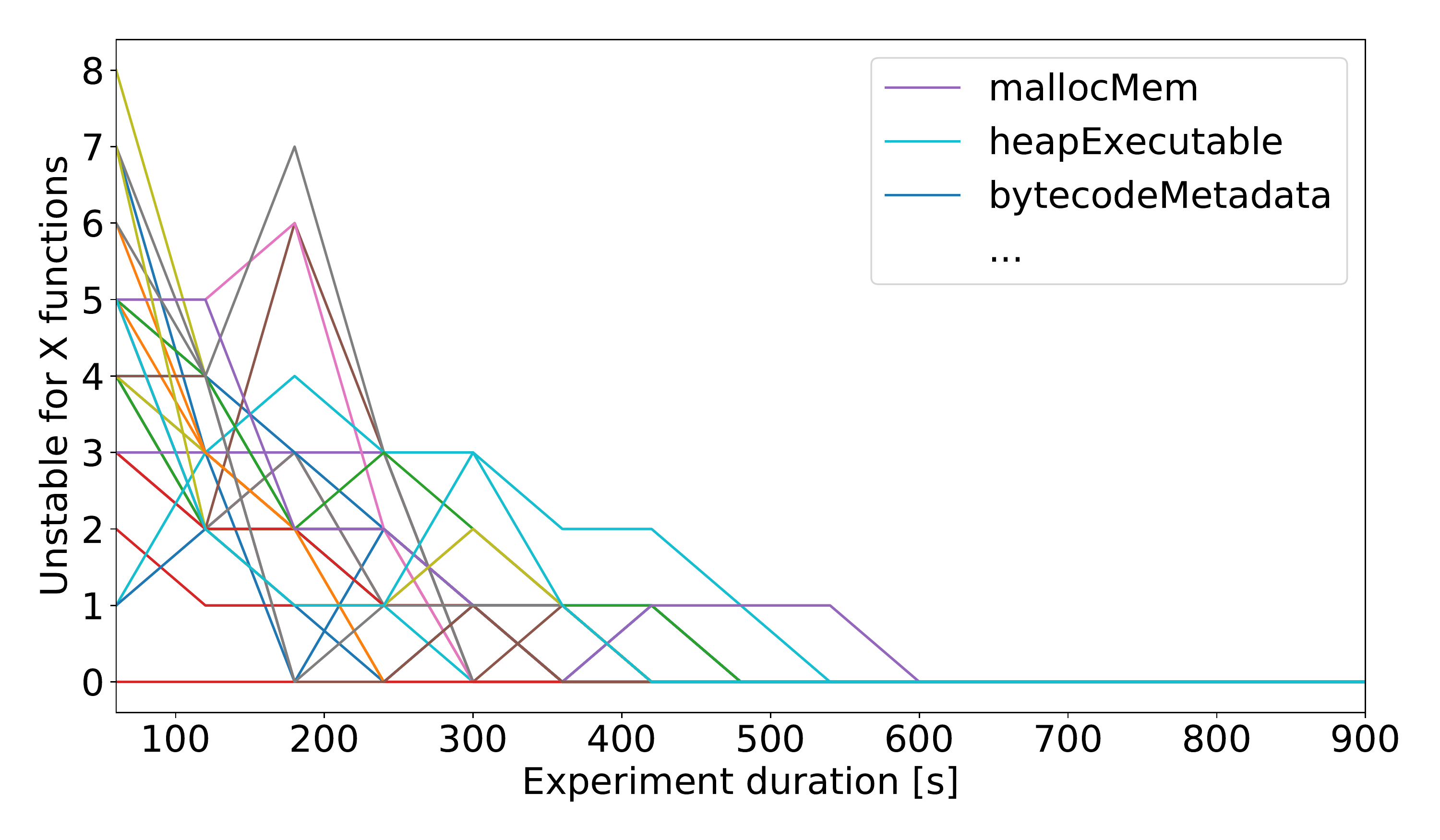}
    \caption{Number of functions for which each metric is unstable for with different measurement duration.}
    \label{fig:DurationImpact}
\end{figure}

\subsection{Resource consumption monitoring}
\label{a2}
We propose to predict the execution time of serverless functions for different memory sizes based on the resource consumption metrics and execution time for a single memory size. However, Lambda currently does not support monitoring of resource consumption metrics out of the box. In general, Lambda's monitoring capabilities are quite lacking, which spawned several third-party monitoring solutions (e.g., by Epsagon, Datadog, and Dynatrace). However, these third-party monitoring solutions focus on tracing requests across a serverless application. If they include resource consumption metrics, they are limited to basic metrics such as CPU utilization. 

Therefore, we implement a custom resource consumption monitoring to cover a wide variety of resource consumption metrics. Table~\ref{tbl:metrics} gives an overview of the metrics our resource consumption monitoring can collect and how the metrics are obtained. The execution time is monitored by timing the execution of the monitored function. A set of resource consumption metrics can be obtained via the Node.js library \texttt{`Process'}. This includes information about the CPU time consumed by user and system processes, the number of voluntary and involuntary context switches, the number of times the file system had to perform I/O, information about the resident set, and the heap usage. Additional information about the heap is collected from the underlying V8 JavaScript engine. Information about the bytes and packages received and transmitted via network is read using Linux counters in \texttt{/proc/net/dev/}. Finally, the Node.js event loop is monitored via the \texttt{perf\_hooks} library. 
Within a traditional VM or container, many additional resource consumption metrics could be collected. However, many of these metrics are either unavailable within serverless functions, always return zero, or return unrealistic values~\cite{PeekingBehindTheCurtains}.

To implement the resource consumption monitoring, we employ a wrapper-style approach, where the monitoring itself implements the Lambda entry point. Whenever the Lambda is triggered, we log the initial metric values for all monitored metrics. Next, the entry point of the original monitored Lambda is called, resulting in a normal execution of this function. After the monitored function is finished, the metrics are polled again and the difference between before and after the function execution is calculated.  The resulting metric values are written to a DynamoDB table. This call to DynamoDB does not affect the collected resource consumption metrics, as it occurs after the metric collection is finished. Finally, the response from the monitored function is returned. 

This wrapper-style approach has been the best practice to monitor Lambda functions since the inception of this service~\cite{Yan2020Extensions}, even though it creates a slight performance overhead. However, note that this overhead does not impact the execution time measurements since we only measure the inner function execution. Recently, AWS previewed Lambda Extensions, which are processes that can run concurrently to the Lambda execution, similar to sidecars for containers~\cite{Extensions}. Lambda Extensions are mostly targeted at making the monitoring of Lambda functions more efficient, and accordingly, we will reimplement our resource consumption monitoring as a Lambda extension once they reach general availability.

\subsection{Dataset generation}
In this work, we conducted extensive performance experiments to create a large dataset describing how different functions are impacted by different memory sizes. Towards this goal, we randomly generated 2\,000 different synthetic functions using our synthetic function generator and equipped them with our resource consumption monitoring. To manage the large number of required performance measurements, we implemented a fully automated measurement harness in Go that relies on Vegeta\footnote{\url{https://github.com/tsenart/vegeta}} as a load driver and enables parallelization of the experiments. For the implementation details of the measurement harness, we refer to our replication package\footnoteref{r1}.

Before we generate the dataset, we need to determine how long each performance experiment needs to run until the reported metrics are stable. To investigate this, we generated 50 functions and measured their execution time and resource consumption metrics for fifteen minutes at 30 requests per second. Next, we tested for each collected metric if the samples across all requests within the first minute, first two minutes, first three minutes, and so on come from the same distribution as the values collected during the full experiment. Figure~\ref{fig:DurationImpact} shows the results when using the Mann–Whitney U test~\cite{mann1947} to test for similarity. We can see, that even after one minute, all metrics are already stable for over 80\% of the investigated functions. If we apply Cliff's delta~\cite{cliff1993dominance} for the differences observed after one minute, all differences are already considered negligible. After ten minutes of experiment time, \texttt{mallocMem} is the last metric to become stable for all functions according to the Mann-Whitney U test. We select ten minutes as the experiment duration for the dataset generation to ensure that stable metrics are collected. 

Finally, we used the measurement harness to measure the execution time and resource consumption metrics for 2\,000 functions across six different memory sizes (128MB, 256MB, 512MB, 1024MB, 2048MB, 3008MB), including the smallest and largest available memory sizes on AWS for ten minutes each at 30 requests per second with an exponentially distributed inter-arrival time. This amounts to 12\,000 performance measurements, 120\,000 minutes of experiment time, 216\,000\,000 Lambda executions, and roughly \$2\,000 worth of Lambda compute time. %
The resulting dataset is publicly available as part of our replication package\footnoteref{r1}. In the following, we describe how we use this dataset to train a multi-target regression model to predict the execution time of a serverless function across all memory sizes based on monitoring data of a single memory size.
\label{a3}

\subsection{Multi-target regression modeling}
\label{sec:modeling}
We formulate the task of predicting the execution time of a serverless function across all memory sizes based on monitoring data from a single memory size as a multi-target regression problem. Hence, for each monitored memory size (base memory size) we train a regression model that predicts the remaining five memory sizes (target memory sizes) based on the average monitored execution time and average resource consumption metrics. To equalize the scale of the target variables, we introduce a preprocessing step that expresses all target execution times as ratios of the input execution time. 

\begin{figure}[t]
    \centering
    \includegraphics[width=\linewidth]{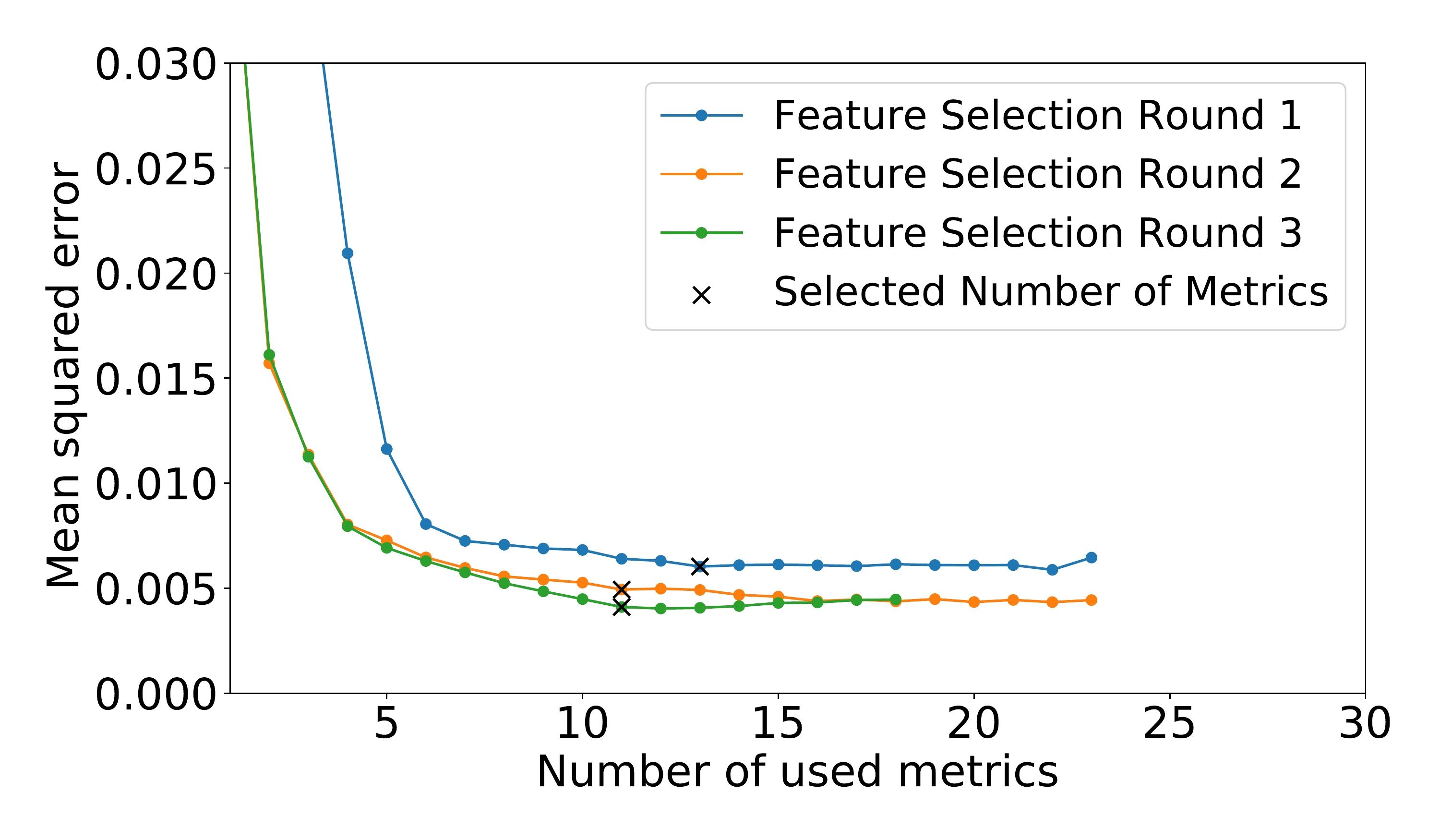}
    \caption{Accuracy and selected metrics for three sequential forward feature selection rounds.}
    \label{fig:featureselecion}
\end{figure}

\begin{figure*}
    \centering
    \begin{subfigure}[b]{0.31\textwidth}
        \centering
        \includegraphics[width=\textwidth]{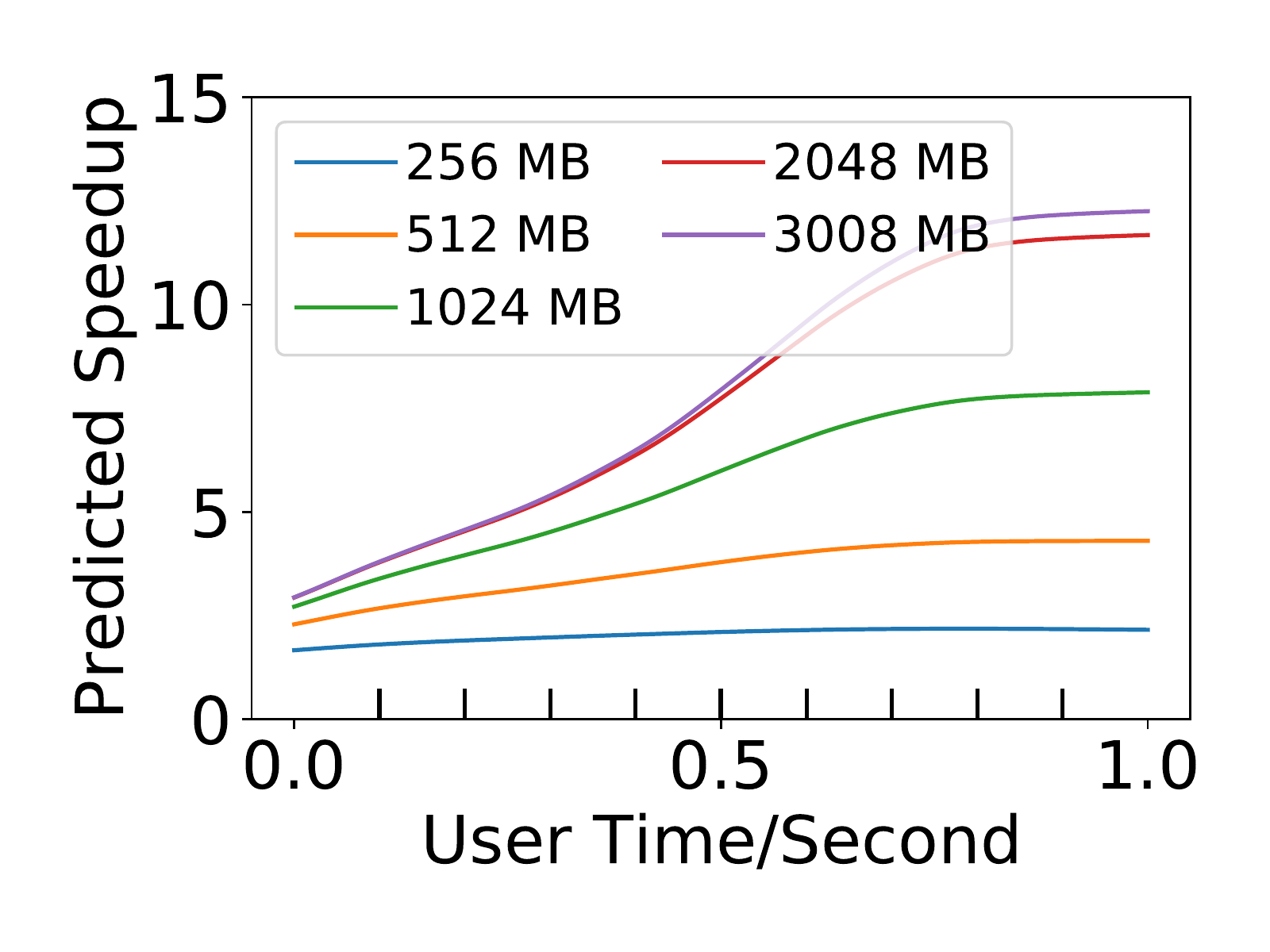}
    \end{subfigure}
    \begin{subfigure}[b]{0.31\textwidth}   
        \centering 
        \includegraphics[width=\textwidth]{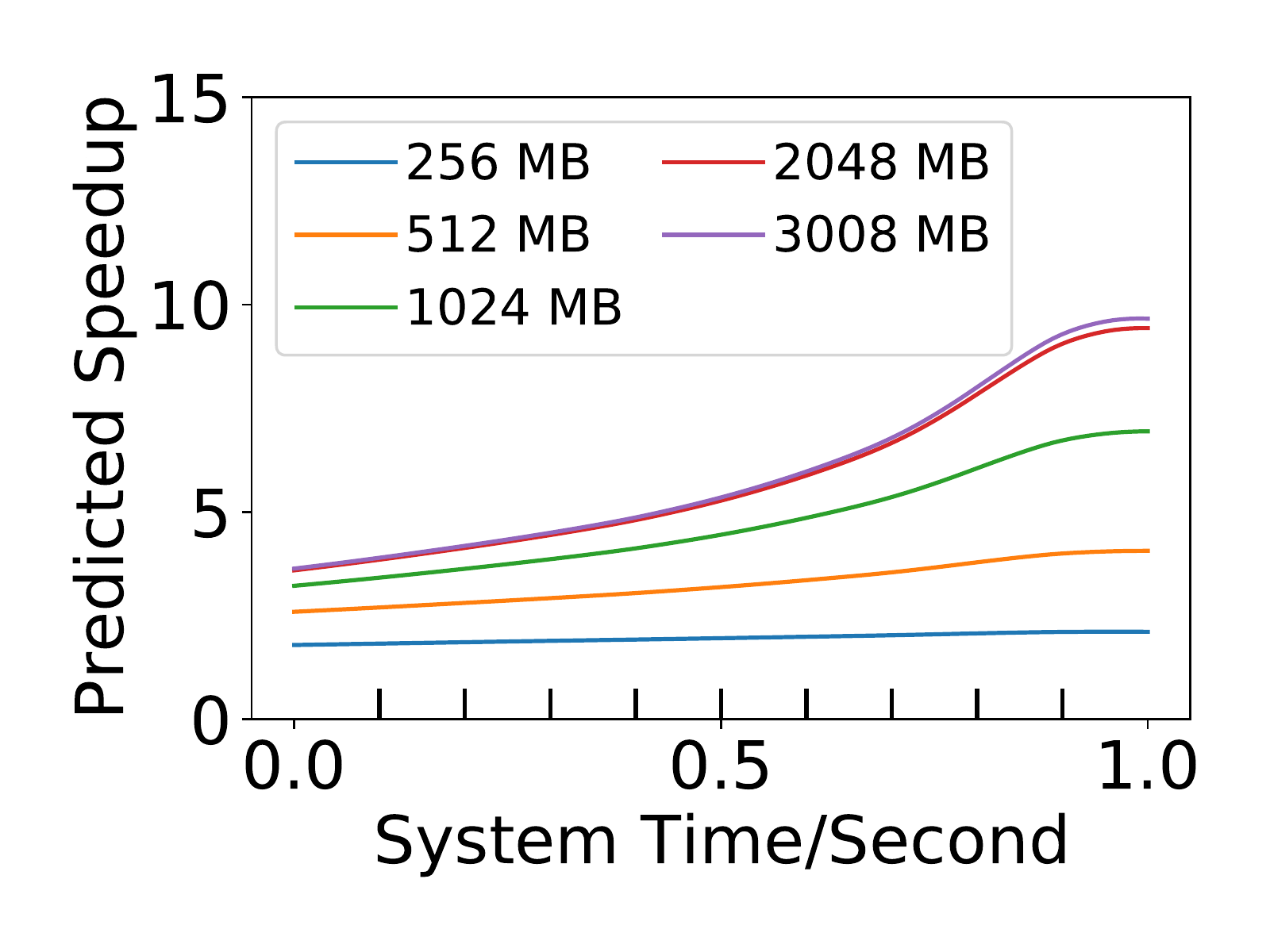}
    \end{subfigure}
    \begin{subfigure}[b]{0.31\textwidth}   
        \centering 
        \includegraphics[width=\textwidth]{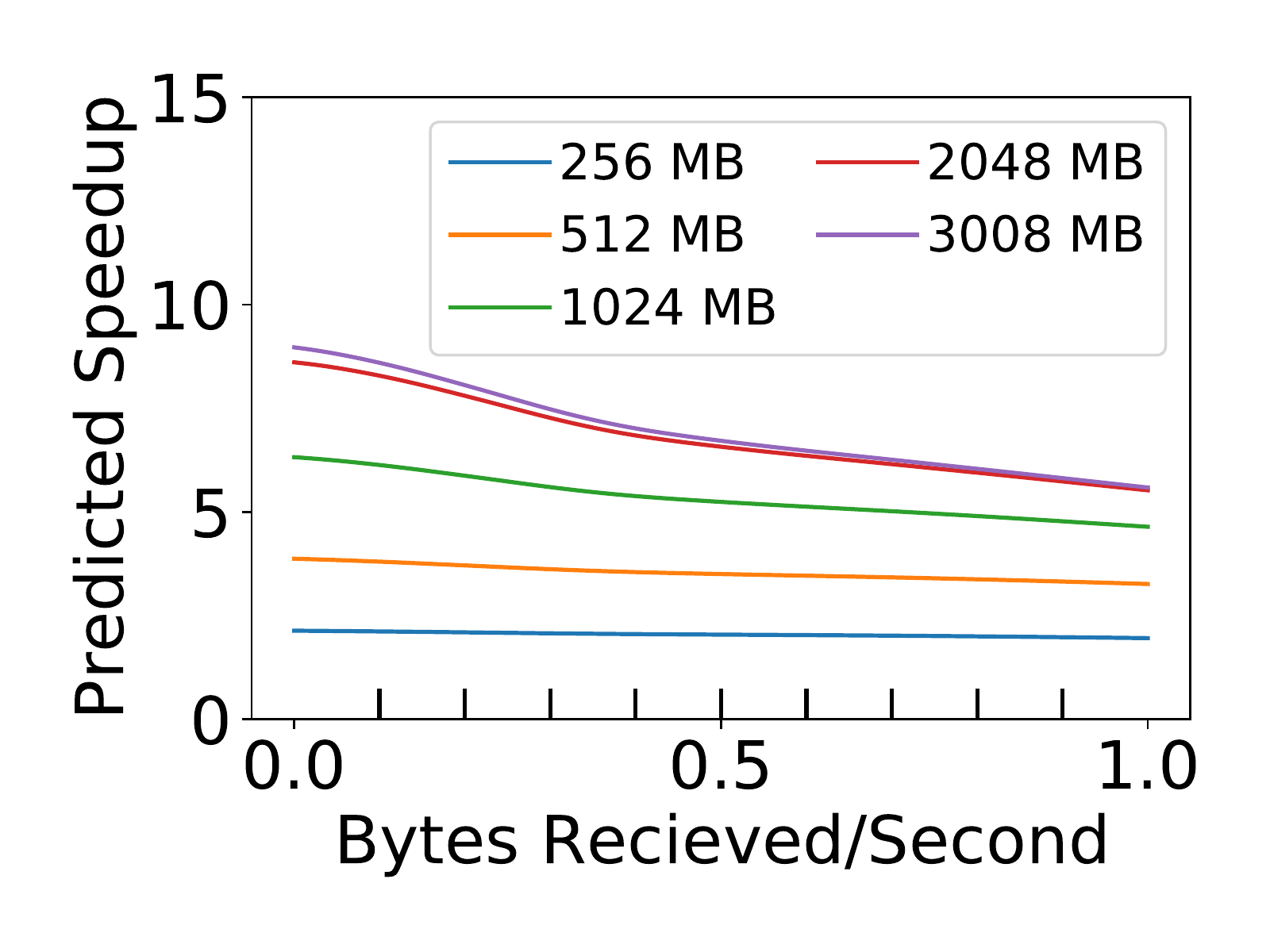}
    \end{subfigure}
    \begin{subfigure}[b]{0.31\textwidth}   
        \centering 
        \includegraphics[width=\textwidth]{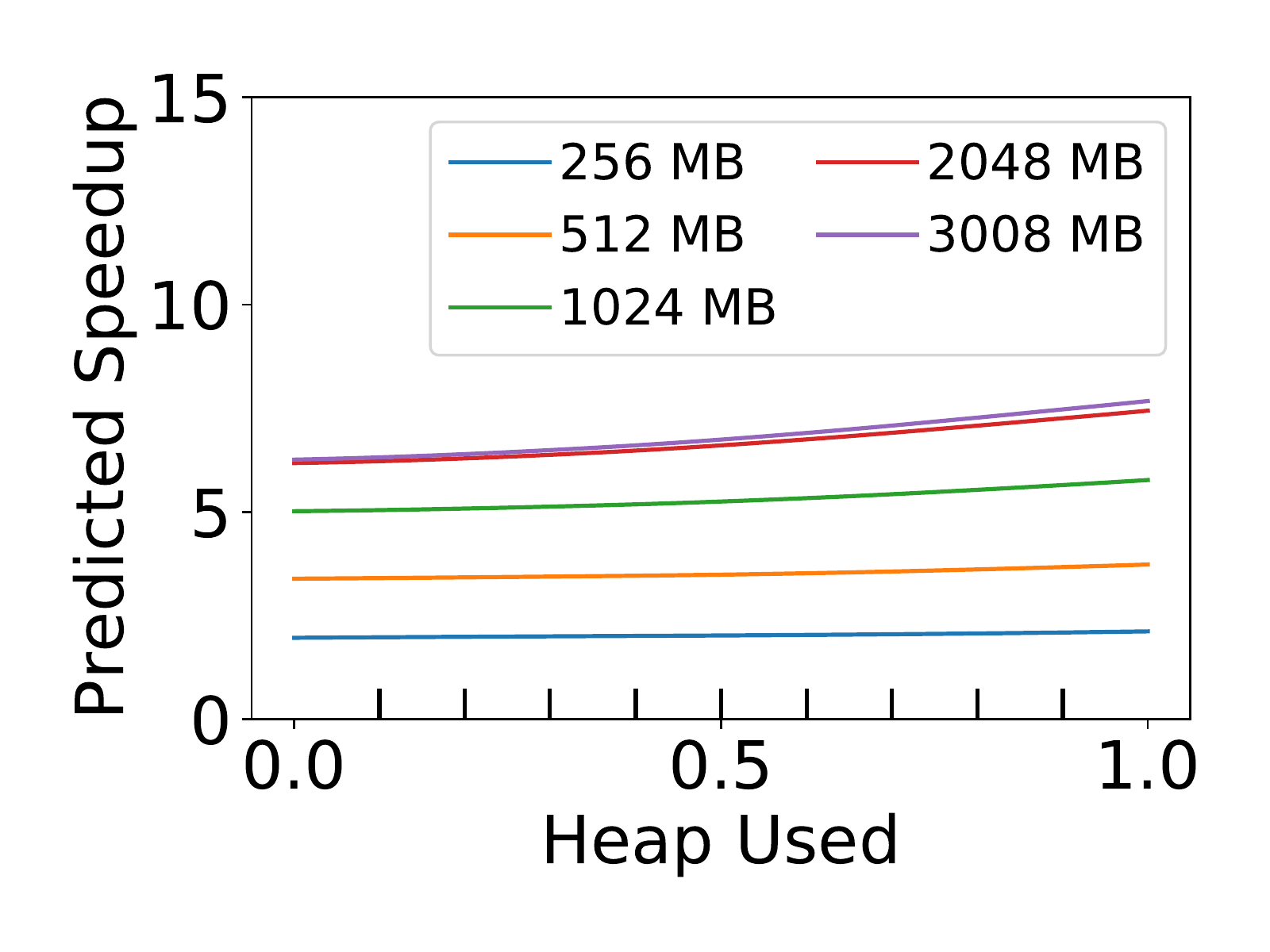}
    \end{subfigure}
    \begin{subfigure}[b]{0.31\textwidth}   
        \centering 
        \includegraphics[width=\textwidth]{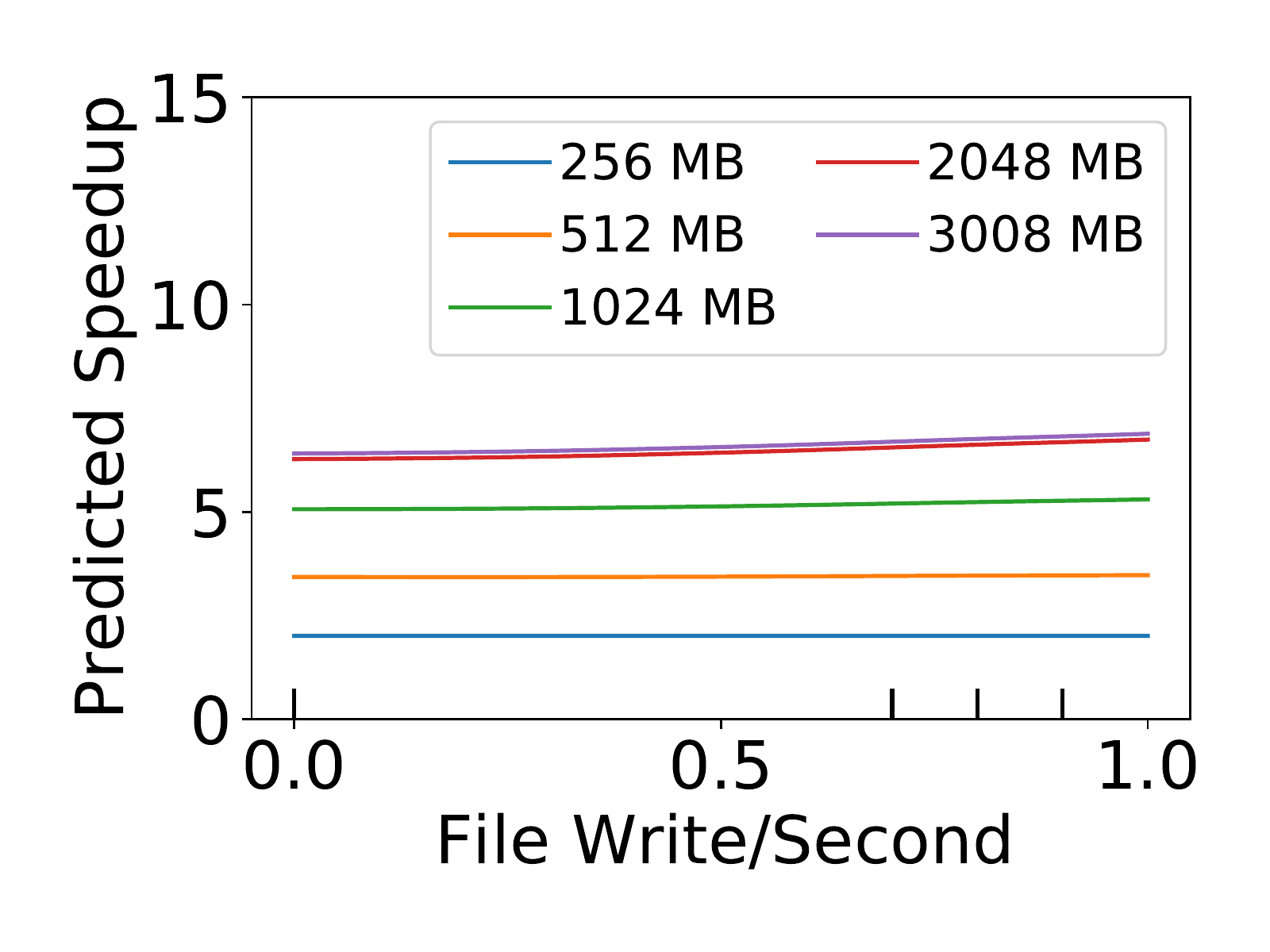}
    \end{subfigure}
    \begin{subfigure}[b]{0.31\textwidth}  
        \centering 
        \includegraphics[width=\textwidth]{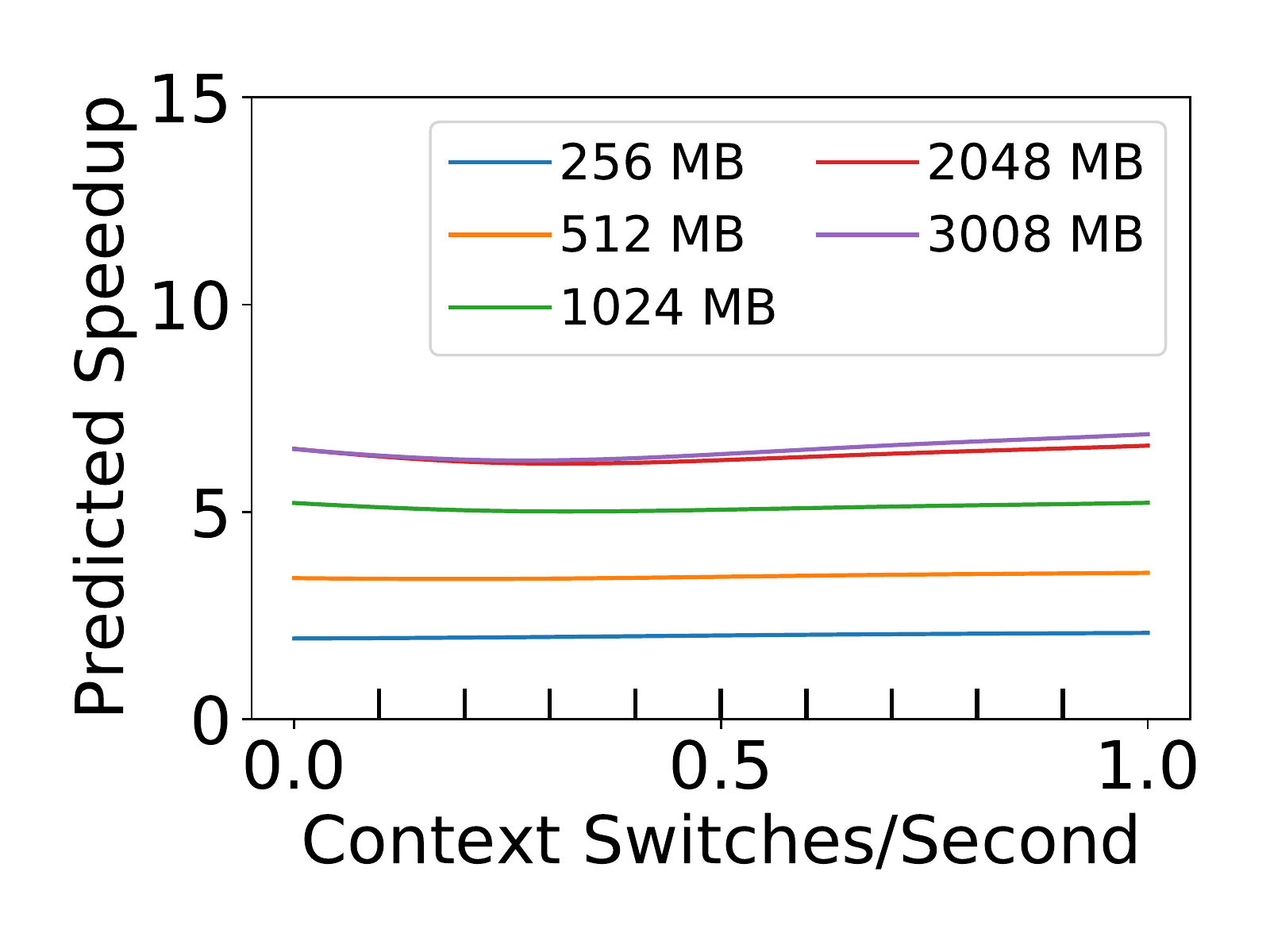}
    \end{subfigure}\caption{Partial dependence plots of the model with basesize 128MB, showcasing the impact of six most important input features on the predicted speedup for different memory sizes. X-axis values are scaled to [0,1]. The partial dependence plots for the remaining features can be found in our replication package.} 
    \label{fig:explain}
\end{figure*}

We start out with a simple neural network with three layers of 128 neurons trained for 200 epochs. Using this model, we conduct several feature engineering and selection rounds, inspired by~\cite{Grohmann2019Monitorless}. The first round of feature selection uses our initial features $F_0$, i.e., the mean execution time and the mean of each resource consumption metric. Figure~\ref{fig:featureselecion} shows that the accuracy increases until we reach thirteen features, so we discard all remaining metrics for the reduced feature set $F_1$. Next, we construct relative features that normalize the $F_1$ features by execution length to obtain $F_2$. For example, in addition to the total number of context switches, $F_2$ also contains the context switches per second. Running the sequential forward feature selection again (see Figure~\ref{fig:featureselecion}) shows an increased model accuracy. We again decrease the number of features in $F_3$ by selecting only the eleven most promising features from $F_2$. Finally, for each remaining metric in $F_3$, we add the standard deviation and coefficient of variation and run the feature selection for the third time to receive the final feature set $F_4$. This results in only a slight accuracy increase, but further reduces the number of base metrics required as all eleven final metrics in $F_4$ are calculated using the metrics heap used, user CPU time, system CPU time, voluntary context switches, bytes written to file system, and bytes received over network. Therefore, our approach requires monitoring only these six metrics when applied in practice.

In order to explore the inner workings of our model and therefore to understand how different characteristics influence the scaling of a function with additional resources, we employ partial dependence plots. A partial dependence plot shows the marginal effect of a feature on the model prediction, which makes it a common explainability tool for machine learning models~\cite{goldstein2015peeking}. Figure~\ref{fig:explain} shows the partial dependence plots for the six most impactful features of our model for a base size of 128MB. It shows that the relative time spent in user and system space (i.e., the CPU utilization caused by the function) have the largest impact on the scaling behavior, with a higher CPU utilization resulting in higher expected speedups for an increased memorysize. This is in line with current assumptions that CPU-intensive functions benefit the most from larger memory sizes~\cite{Casalboni2020Deep}. Interestingly, the number of bytes received per second correlates negatively with the predicted speedup, so a function that is network-intensive will scale worse with larger memory sizes. Further, the heap used (so the memory used by the application) also has an impact on the predicted speedup. If a function uses a lot of its available memory, adding additional memory would reduce memory swapping, whereas for a function that already has sufficient memory available additional memory would not be beneficial. While the remaining features also impact the final prediction, they seem to be mostly used to fine-tune the prediction. To summarize, we find that the predicted speedup mostly depends on the CPU utilization, network activity and memory used by the function.

Next, we conduct a grid search to tune the hyperparameters of the model. Table~\ref{tab:hyperparameter} shows the tuned parameters, the ranges for each parameter, and the parameter value selected by the grid search. The final model uses the Adam optimizer, a MAPE loss function, 200 epochs, an L2 regularization of $10^{-2}$, and four layers. The training of this model takes about three minutes, so the training overhead is negligible.

As our approach relies on monitoring data from the function running in production with a single memory size, we investigate if a certain monitored memory size provides better prediction accuracy than others, as it would then make sense to deploy the function with this memory size initially.
Therefore, we run ten iterations of five-fold cross-validation with a random split for each base memory size. Table~\ref{tab:basesize} shows the resulting mean squared error, mean absolute percentage error, coefficient of determination ($R^2$), and explained variance score, which are common metrics to determine the performance of regression models~\cite{glantz2001primer}. We select 256MB as the default basesize, as it shows the best mean squared error, the second-highest R\textasciicircum{}2 and explained variance scores, and a good mean absolute percentage error.

In the following, we describe how the execution time prediction of this model can be used to automatically optimize the memory size of a serverless function.

\begin{table}[t]
\centering
\caption{Parameter range and selected parameters for the hyperparameter optimization.}
\label{tab:hyperparameter}
\begin{tabular}{@{}lll@{}}
\toprule
Parameter & Parameter range & Selected \\ \midrule
Optimizer & SGD, Adam, Adagrad & Adam \\
Loss & MSE, MAE, MAPE & MAPE \\
Epochs & 200, 500, 1000 & 200 \\
Neurons & 64, 128, 256 & 256 \\
L2 & 0, 0.0001, 0.001, 0.01 & 0.01 \\
Layers & 2, 3, 4, 5 & 4 \\ \bottomrule
\end{tabular}
\end{table}

\begin{table}[t]
\caption{MSE, MAPE, R\textasciicircum{}2, and explained variance for each base memory size (in MB) based on cross-validation.}
\label{tab:basesize}
\begin{tabular}{@{}lcccccc@{}}
\toprule
Basesize & 128 & \textbf{256} & 512 & 1024 & 2048 & 3008 \\ \midrule
MSE & 0.005 & \textbf{0.003} & 0.004 & 0.009 & 0.010 & 0.015 \\
MAPE & 0.066 & \textbf{0.046} & 0.040 & 0.031 & 0.033 & 0.036 \\
R\textasciicircum{}2 & 0.986 & \textbf{0.977} & 0.971 & 0.970 & 0.954 & 0.958 \\
ExpVar & 0.987 & \textbf{0.979} & 0.974 & 0.972 & 0.962 & 0.963 \\ \bottomrule
\end{tabular}
\end{table}

\subsection{Memory size optimization}
\label{sec:optimization}
Automatically determining the optimal memory size for a serverless function results in a standard multi-objective optimization problem, as we want to optimize for both performance and cost. A common approach to determine a single, optimal solution for multi-objective optimization problems is to use a parameterizable tradeoff function that combines the objectives into a single score~\cite{rao2019engineering}. This requires scores of the same scale for each objective, which we calculate for each memory size $m_x$ as following:

\begin{equation*}
    S_{cost}(m_x) =\frac{\text{cost}(m_x)}{
    \min_{\forall m_i \in M} \text{cost}(m_i)},
\end{equation*}
\begin{equation*}
    S_{perf}(m_x) =\frac{\text{executionTime}(m_x)}{
    \min_{\forall m_i \in M} \text{executionTime}(m_i)}
\end{equation*}
with $M$ being the set of all available memory sizes and the functions $ \text{cost}()$ and $\text{executionTime}()$ returning the predicted cost and execution time for a given memory size $m_i$. Note that for monitored memory sizes the observed values can be used. Both scores have a minimum of 1, which indicates an optimal cost/execution time, and each value above one indicates the percentage deviation from the optimum, so a $S_{cost}$ of 1.5 would indicate a 50\% increased cost compared to the lowest possible cost for this function. Therefore, both scores use the same scale and are humanly interpretable. For the final objective function, these scores are combined using a configurable tradeoff value $t \in [0,1]$ as follows:

\begin{equation*}
    S_{total}(m_x) = t \cdot S_{cost}(m_x) + (1 - t) \cdot S_{perf}(m_x)
\end{equation*}

The value for $t$ can be set by the system operator based on their preferences, where $t = 0.5$ would indicate that a one percent increase in cost compared to the best possible cost is worth the same as a one percent increase in execution time compared to the best possible execution time is worth the same. 
Whereas a value of $t = 0.75$ would result in accepting an X\% increase in cost only if it would result in a 3X\% decrease of execution time. 

In order to apply this tradeoff function to optimize the memory size of serverless functions, we first use our model to predict the execution time for all memory sizes and then calculate the predicted cost for a function execution based on the pricing model of the cloud provider, as it only depends on the execution time. Next, we calculate $S_{total}$ for each memory size and select the memory size with the lowest $S_{total}$ score:

\begin{equation*}
    OptSize = \argmin_{\forall m_x \in \{128, 256, 512, 1024, 2048, 3008\}} S_{total}(m_x)
\end{equation*}

\section{Evaluation}
\label{Evaluation}

\begin{figure*}[t!]
    \centering
    \begin{subfigure}[b]{0.244\textwidth}
        \centering
        \includegraphics[width=\textwidth]{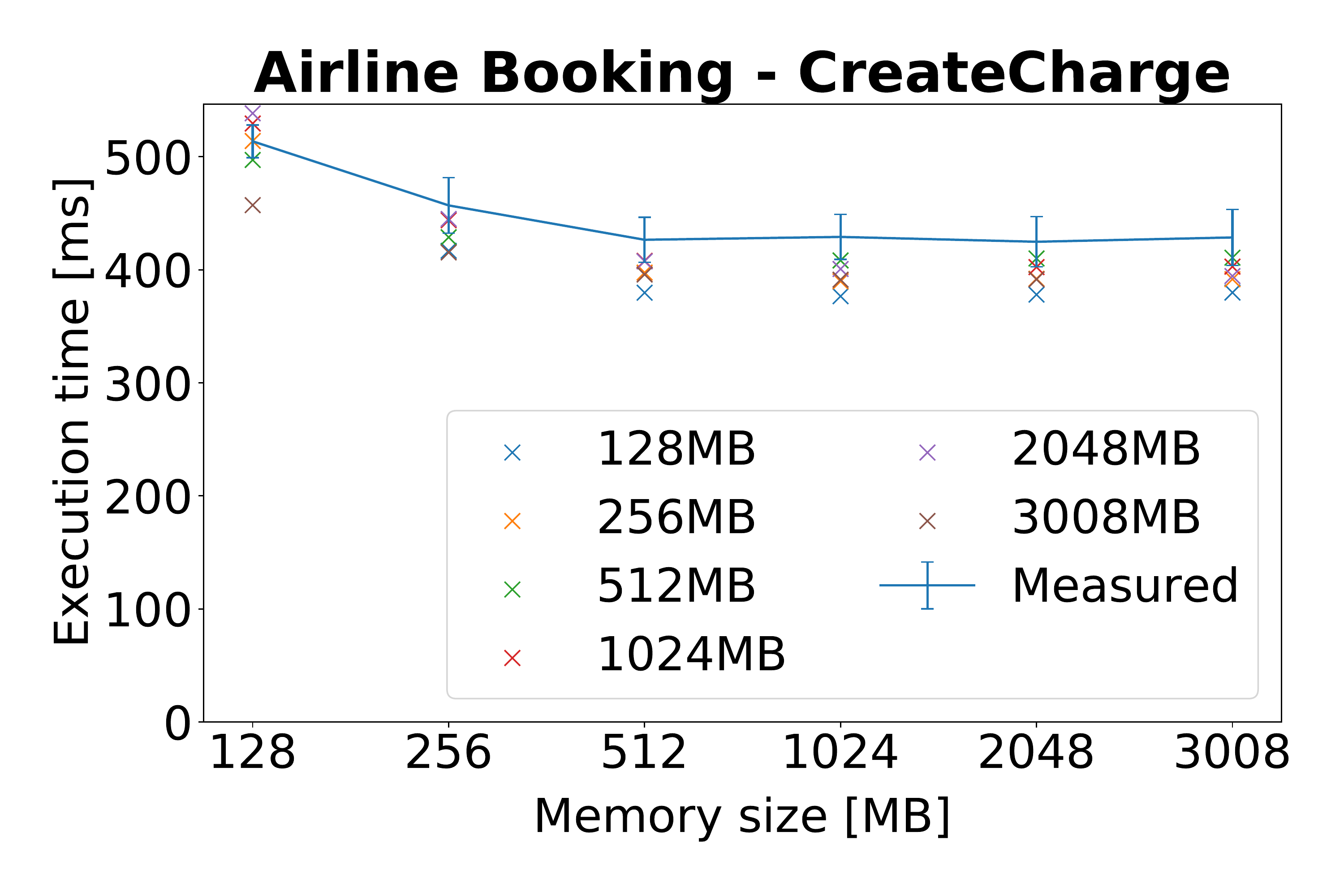}
    \end{subfigure}
    \hfill
    \begin{subfigure}[b]{0.244\textwidth}  
        \centering 
        \includegraphics[width=\textwidth]{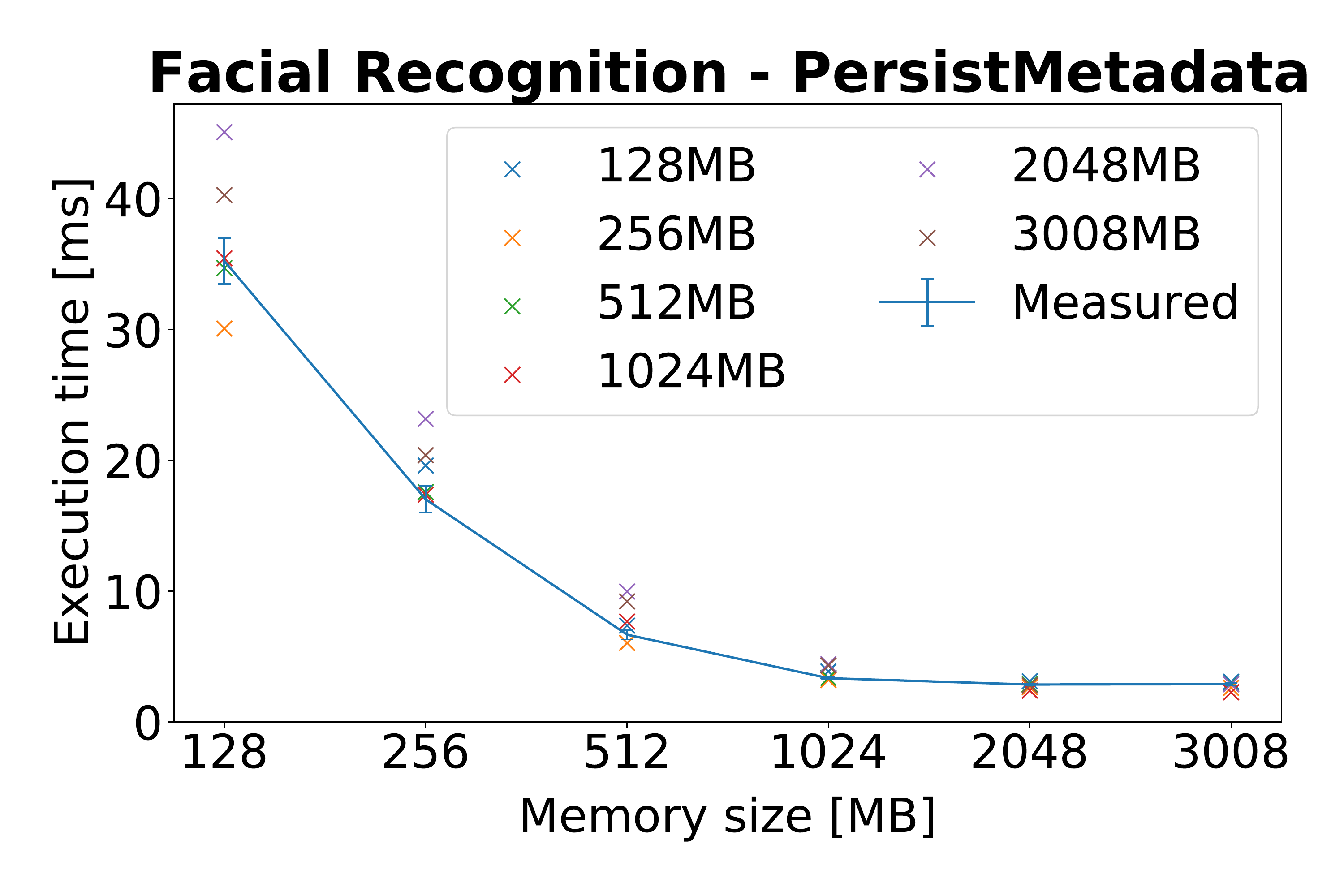}
    \end{subfigure}
    \begin{subfigure}[b]{0.244\textwidth}  
        \centering 
        \includegraphics[width=\textwidth]{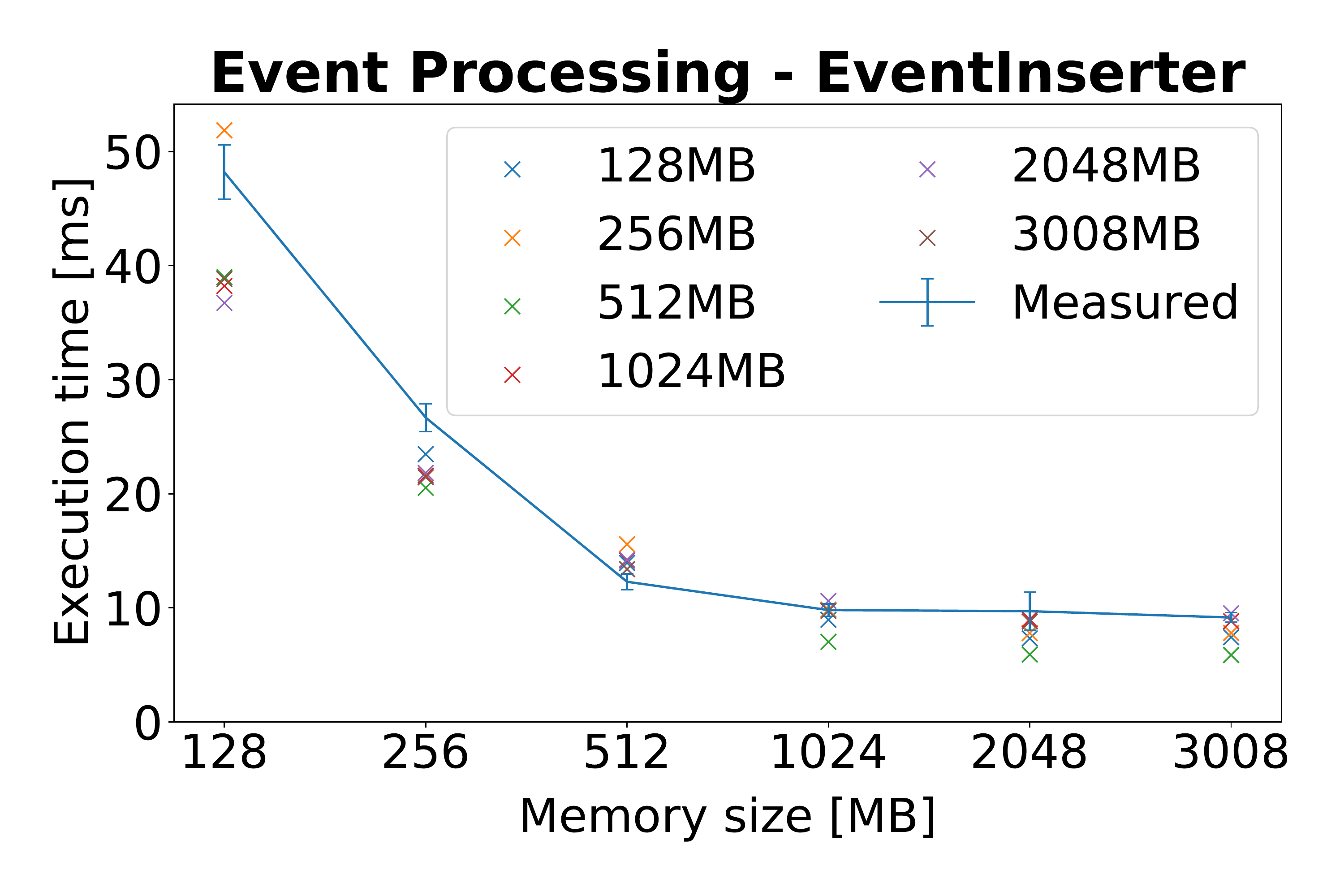}
    \end{subfigure}
    \begin{subfigure}[b]{0.244\textwidth}  
        \centering 
        \includegraphics[width=\textwidth]{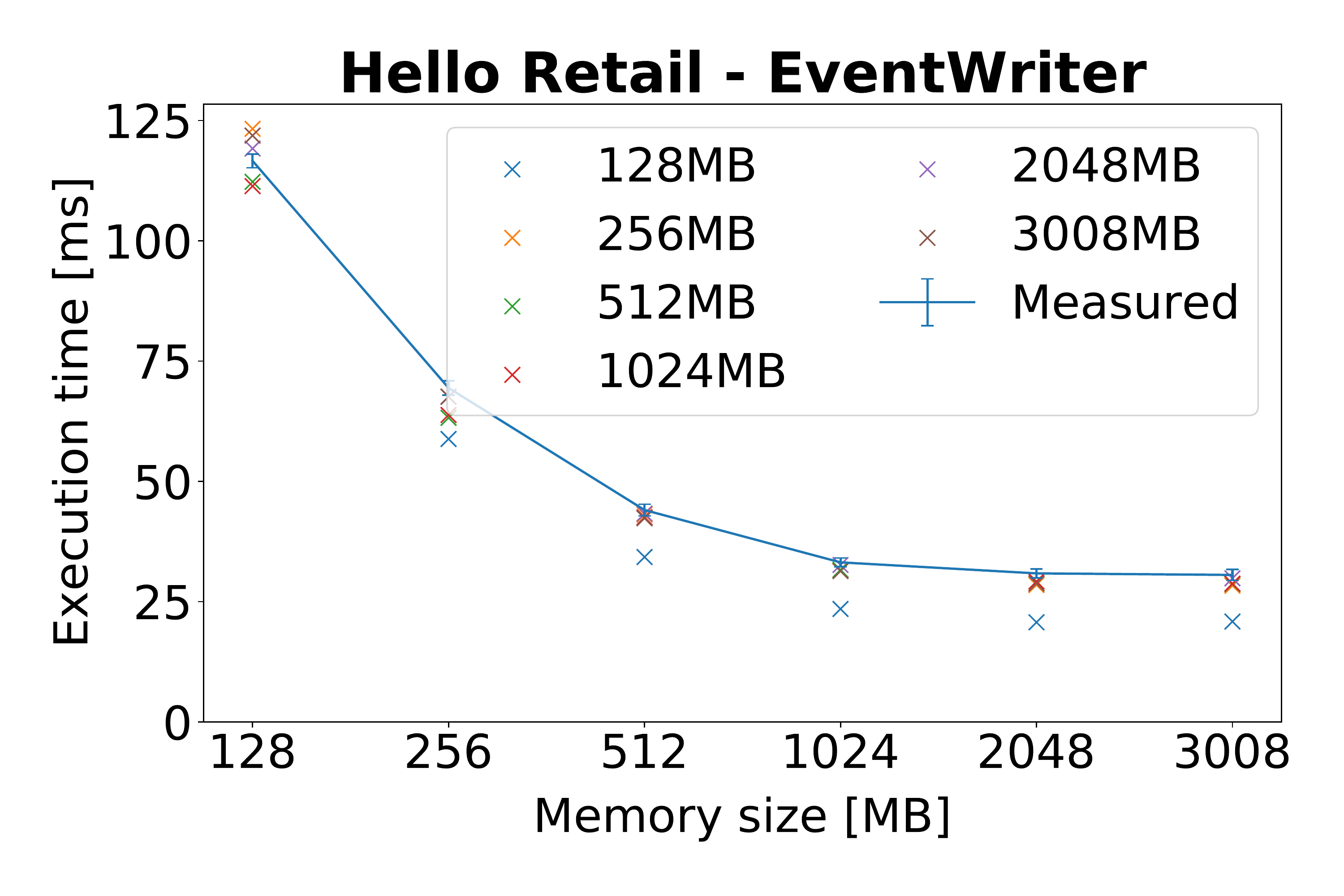}
    \end{subfigure}
    \begin{subfigure}[b]{0.244\textwidth}  
        \centering 
        \includegraphics[width=\textwidth]{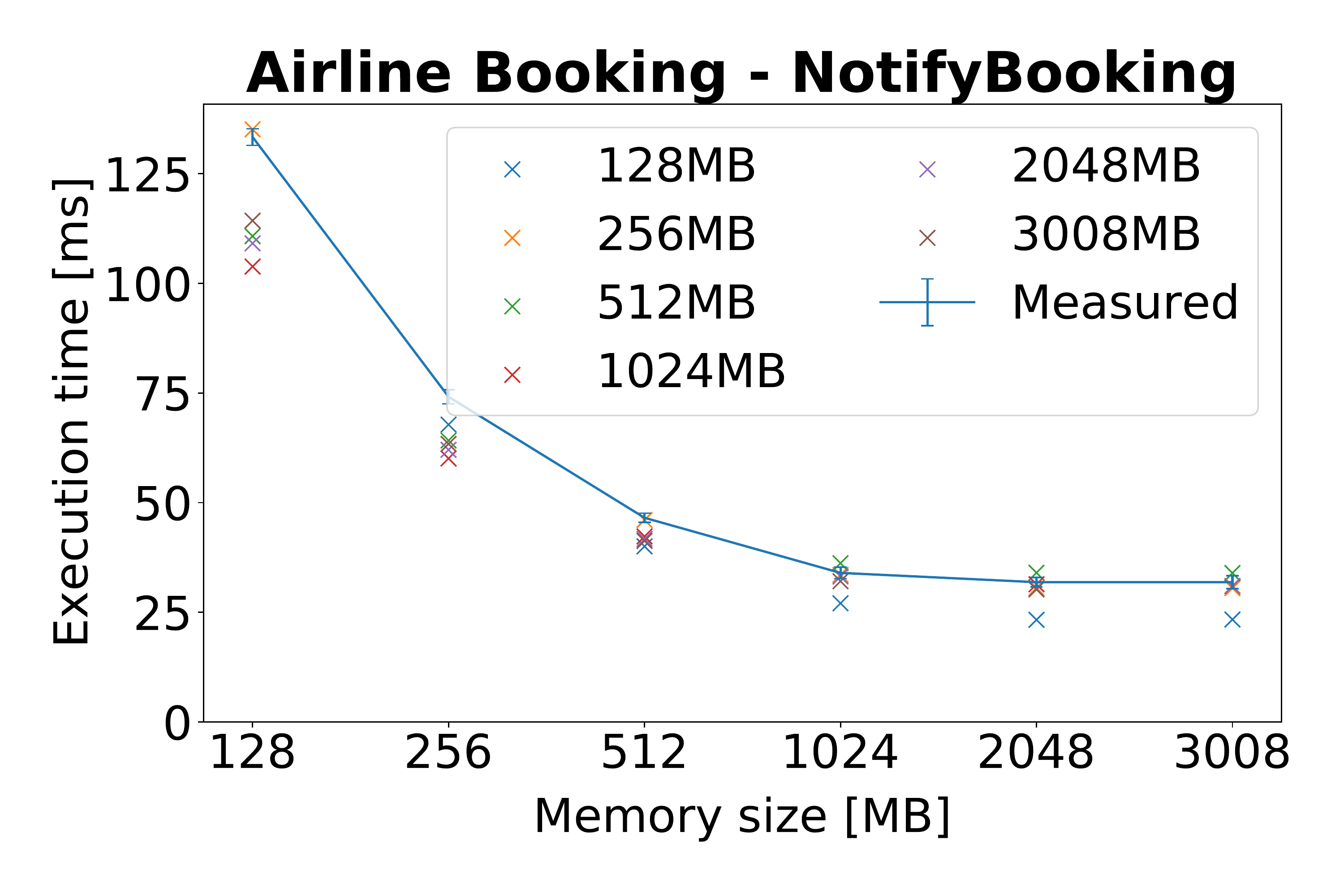}
    \end{subfigure}
    \begin{subfigure}[b]{0.244\textwidth}
        \centering
        \includegraphics[width=\textwidth]{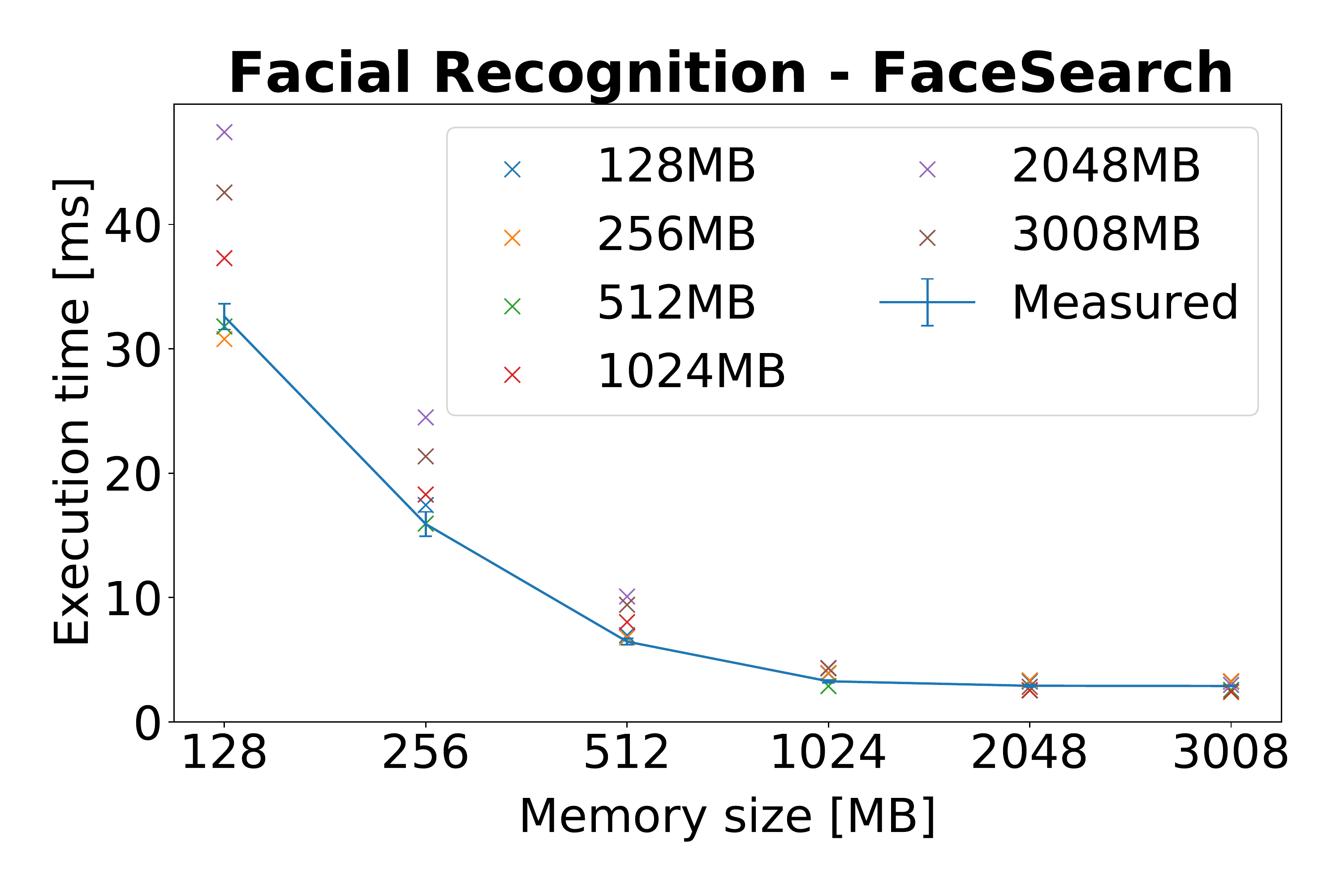}
    \end{subfigure}
    \hfill
    \begin{subfigure}[b]{0.244\textwidth}  
        \centering 
        \includegraphics[width=\textwidth]{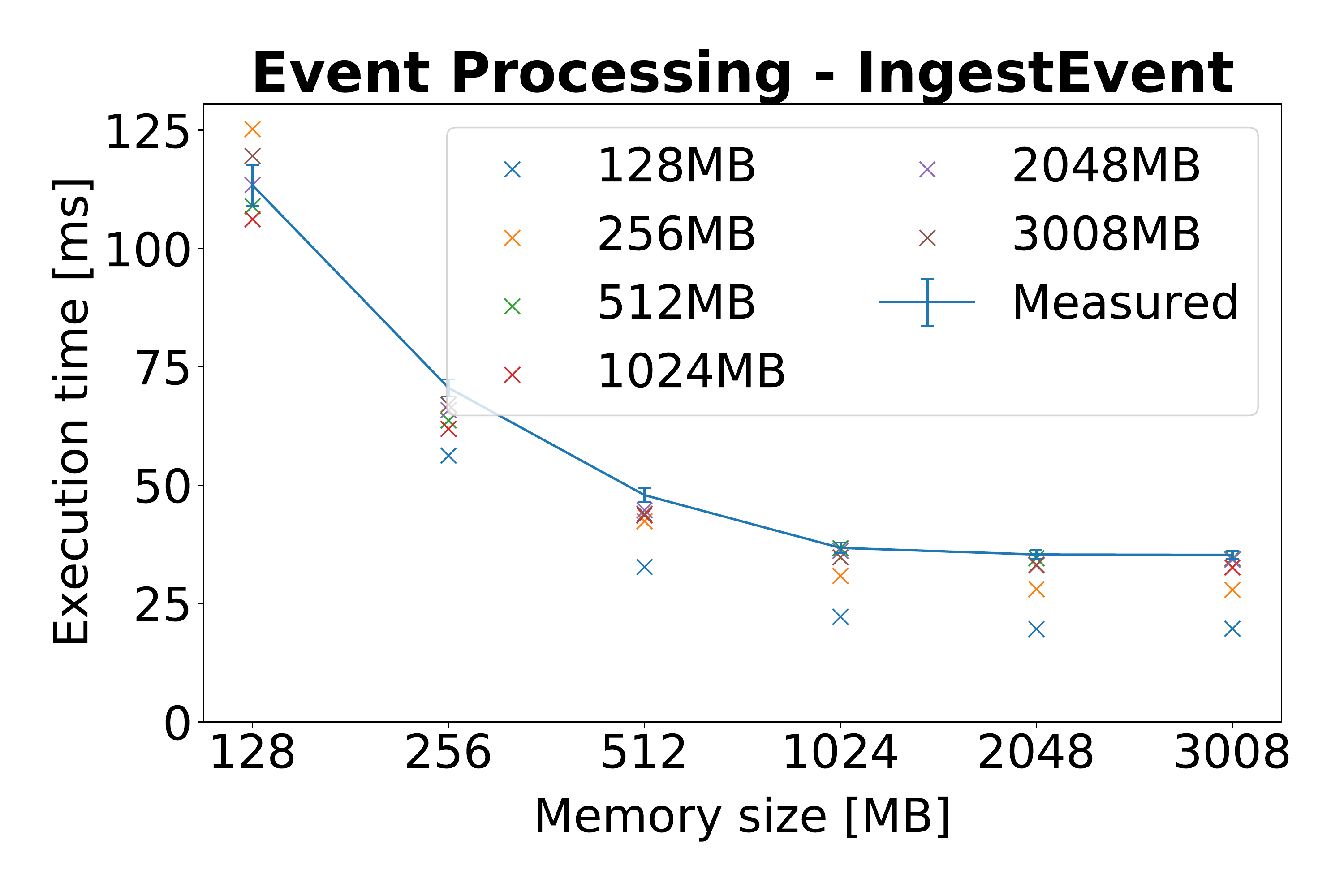}
    \end{subfigure}
    \begin{subfigure}[b]{0.244\textwidth}  
        \centering 
        \includegraphics[width=\textwidth]{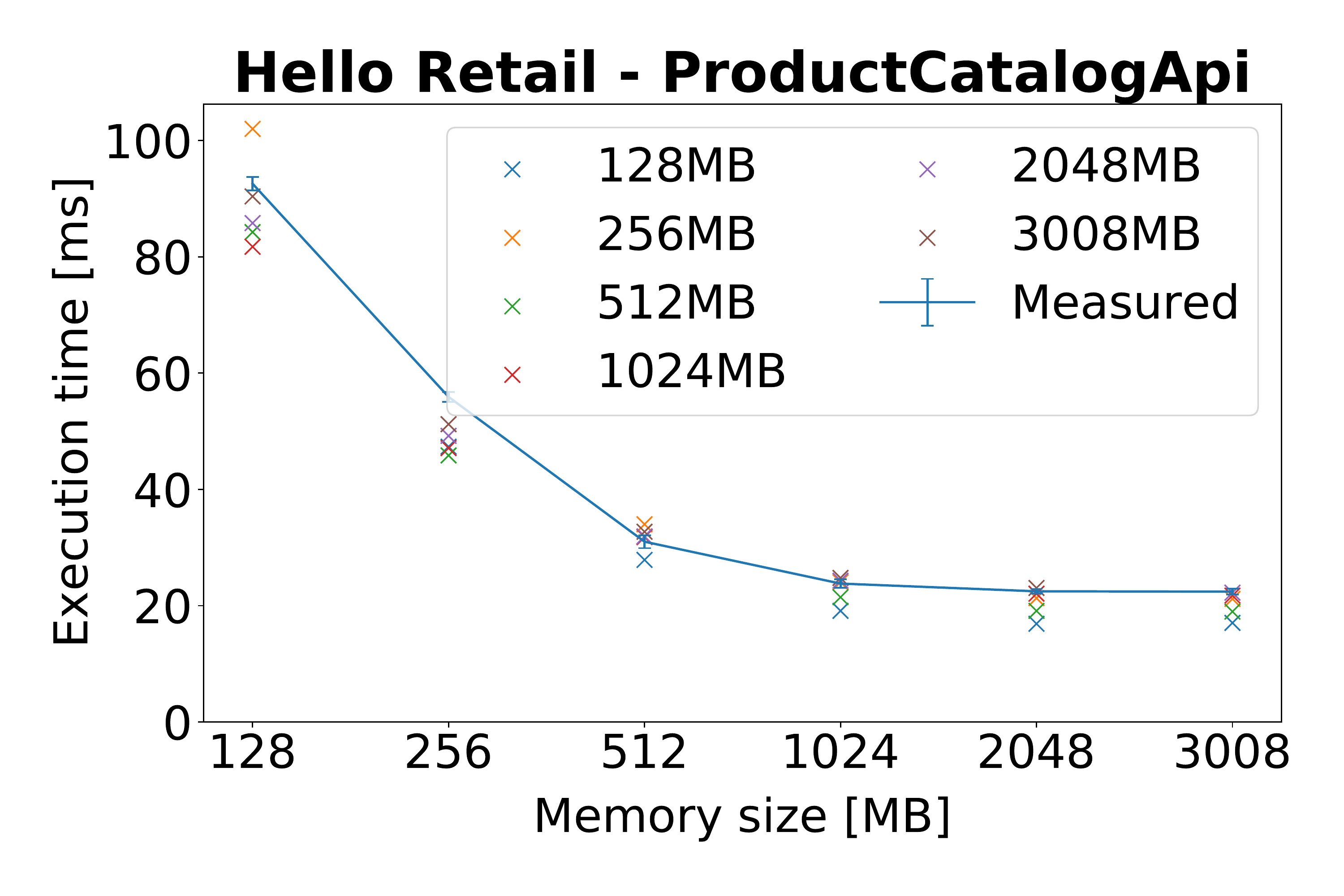}
    \end{subfigure}
    \caption{Example for the measured and predicted execution time for a serverless function of each serverless application.} 
    \label{fig:airlinePrediction}
\end{figure*}

We design our evaluation to answer the following three research questions:
\begin{itemize}
    \item \textbf{RQ1:} \textit{Can our model, trained on a synthetic dataset, accurately predict the execution time of realistic serverless functions?}
    \item \textbf{RQ2:} \textit{Are the execution time predictions provided by our approach sufficient to determine the optimal memory size of serverless functions?}
    \item \textbf{RQ3:} \textit{How large are the benefits in terms of decreased cost and execution time of our proposed approach?}
\end{itemize}
\noindent In order to investigate these research questions, we conduct performance measurements for the following four case study systems: \newline \newline

\vspace{-10px}\noindent\textbf{Airline Booking} This application was the subject of the AWS Build On Serverless series~\cite{buildon} and presented at AWS re:Invent as an example for the implementation of a production-grade full-stack app using AWS Amplify~\cite{reinvent}. The airline booking application implements the flight booking aspect of an airline. Customers can search for flights, book flights, pay using a credit card, and earn loyalty points with each booking. It consists of eight serverless functions, the managed services S3, SNS, Step Functions, and API Gateway, as well as an external payment provider. The workload consists of 200 requests per second that sequentially access all application features for ten minutes. The measurements for this case study were conducted in October 2020, so two months after the training dataset and incurred costs of \char`\~500\$.\newline \newline

\vspace{-15px}\noindent\textbf{Facial Recognition} This case study is part of the AWS Wild Rydes workshop and was also used in the evaluation of Costless~\cite{costless}, another cost optimization approach for serverless functions. In this application, users of a fictional transportation app, Wild Rydes, upload their profile picture, which triggers the execution of a workflow that performs facial recognition, matching, and indexing. It consists of six serverless functions, however, we removed the notification function as it is a no-op stub. This application makes heavy use of AWS Rekognition, a service that was not contained in our function segments. The workload consists of only ten requests per second for five minutes, as AWS Rekognition is comparatively expensive. Therefore, for this case study, our approach will have less monitoring data available to learn from. The measurements for this case study were conducted in December 2020, so four months after the training dataset and incurred costs of \char`\~400\$.\newline \newline

\vspace{-15px}\noindent\textbf{Event Processing} This application was introduced in~\cite{yussupov2019facing}, where the authors investigate the challenges of migrating serverless applications across cloud providers by migrating different systems across multiple cloud providers. For our case study, we use the AWS implementation of the IoT-inspired event processing system, where the data obtained from multiple sensors are aggregated for further processing. It consists of seven serverless functions and uses the API Gateway, SNS, SQS, and AWS Aurora, none of which are used in our function segments. Compared to the other two applications, the functions of this application exhibit very fast execution times. The workload consists of 10 requests per second that sequentially access all application features for ten minutes. The measurements for this case study were conducted in December 2020, so four months after the training dataset and incurred costs of \char`\~50\$.\newline \newline

\vspace{-15px}\noindent\textbf{Hello Retail} This application from the online retailer Nordstrom won the inaugural serverless architecture competition at Serverlessconf Austin~\cite{Nordstrom}. The application models a product catalog, including a workflow for new products that outsources the product image acquisition to photographers. This application consists of seven functions and uses Kinesis, API Gateway, Stepfunctions, DynamoDB, and S3. The workload consists of 10 requests per second that sequentially access all application features for ten minutes. The measurements for this case study were conducted in May 2021, so nine months after the training dataset and incurred costs of \char`\~30\$.

\vspace{5px}\noindent For all four case studies, we conducted ten measurement repetitions for each memory size to account for cloud performance variability, which is in line with recently proposed guidelines for performance measurements of cloud applications~\cite{Papadopoulos2019methodological}. We run the experiments as randomized multiple interleaved trials, which has been shown to reduce cloud performance variability~\cite{Abedi2017conducting}. Additionally, to reduce the chance of human error and enable the replication of our measurement results, we implemented fully automated measurement harnesses for each case study, which are available as part of our replication package\footnote{\label{r2}\url{https://github.com/Sizeless/ReplicationPackage}}. In terms of number of functions and external services used, these four applications are more complex than the average serverless application~\cite{shahrad2020serverless, eismann2020serverless}.\newline \newline

\begin{table}[t]
\centering
\caption{Relative prediction error based on monitoring data from 256MB for the \textbf{airline booking application}.}
\label{tab:airline}
\begin{tabular}{lrrrrr}
\toprule
Targetsize &  128 &  512 &  1024 &  2048 &  3008 \\
\midrule
IngestLoyalty  & 10.2 &  4.7 &  16.8 &  19.6 &  19.8 \\
CaptureCharge  &  0.4 &  7.3 &   8.8 &   7.6 &   8.4 \\
CreateCharge   &  0.1 &  6.8 &   9.2 &   7.7 &   8.6 \\
CollectPayment & 26.1 & 16.8 &  26.7 &  30.3 &  29.2 \\
ConfirmBooking &  5.7 &  9.6 &  11.8 &   6.1 &   5.0 \\
GetLoyalty     & 11.2 & 16.6 &  27.3 &  32.7 &  33.1 \\
NotifyBooking  &  1.3 &  1.1 &   1.3 &   4.1 &   4.3 \\
ReserveBooking &  1.3 & 11.6 &  16.3 &  11.8 &   8.3 \\
\midrule
All functions                    &  7.0 &  9.3 &  14.8 &  15.0 &  14.6 \\
\bottomrule
\end{tabular}
\end{table}

\begin{table}[t]
\centering
\caption{Relative prediction error based on monitoring data from 256MB for the \textbf{facial recognition application}.}
\label{tab:face}
\begin{tabular}{lrrrrr}
\toprule
Targetsize &  128 &  512 &  1024 &  2048 &  3008 \\
\midrule
FaceDetection   &  0.8 &  0.9 &  15.8 &   2.8 &   1.3 \\
FaceSearch      &  5.6 &  5.1 &  20.7 &  15.2 &  13.5 \\
IndexFace       & 12.0 & 18.9 &   9.2 &  17.0 &  18.8 \\
PersistMetadata & 14.7 &  9.5 &   4.3 &   6.9 &   9.7 \\
CreateThumbnail & 30.7 &  6.5 &  25.1 &  10.4 &   6.4 \\
\midrule
All functions                        & 12.7 &  8.2 &  15.0 &  10.5 &   9.9 \\
\bottomrule
\end{tabular}
\end{table}

\vspace{-15px}\noindent\textbf{RQ1:} \textit{Can our model, trained on a synthetic dataset, accurately predict the execution time of realistic serverless functions?}
The underlying idea of our approach is to learn how memory size impacts the execution time of a function based on a large dataset obtained from synthetic functions with realistic resource consumption profiles. To investigate if the knowledge our model learned from the synthetic dataset can be transferred to realistic functions, we train the model on the synthetic dataset as described earlier and ask it to predict the execution time of the twenty-seven serverless functions from our four case study systems across all memory sizes based on monitoring data obtained from a single memory size. Figure~\ref{fig:airlinePrediction} shows the measured execution time and standard deviation across ten measurement repetitions for two functions from each application in blue. The colored crosses show the predictions for different basesizes. The individual functions scale differently with increasing memory sizes, and our approach is generally able to predict these differences quite well. It also shows that there are some inaccuracies, especially when predicting the execution time of 128MB. The graphs for the remaining nineteen functions can be viewed in the CodeOcean capsule of our replication package\footnoteref{r2}.

\begin{table}[t]
\centering
\caption{Relative prediction error based on monitoring data from 256MB for the \textbf{event processing application}.}
\label{tab:event}
\begin{tabular}{lrrrrr}
\toprule
Targetsize &  128 &  512 &  1024 &  2048 &  3008 \\
\midrule
EventInserter   &  7.6 & 26.8 &   0.4 &  19.8 &  14.8 \\
FormatForecast  &  7.7 &  9.1 &   7.1 &   4.6 &   6.3 \\
FormatState     &  8.9 &  2.2 &   6.1 &   5.6 &   8.2 \\
FormatTemp      &  3.1 &  3.4 &   9.6 &   5.5 &   9.7 \\
GetLatestEvents & 23.8 & 56.5 &  70.9 &  50.3 &  47.6 \\
ListAllEvents   & 18.0 & 34.3 & 119.4 & 131.9 & 132.2 \\
IngestEvent     & 10.5 & 11.5 &  16.1 &  20.8 &  20.9 \\
\midrule
All functions                      & 11.4 & 20.5 &  32.8 &  34.1 &  34.2 \\
\bottomrule
\end{tabular}
\end{table}

\begin{table}[t]
\centering
\caption{Relative prediction error based on monitoring data from 256MB for the \textbf{hello retail application}.}
\label{tab:hello}
\begin{tabular}{lrrrrr}
\toprule
Targetsize &  128 &  512 &  1024 &  2048 &  3008 \\
\midrule
EventWriter           &  5.7 &  2.1 &   4.9 &   7.8 &   7.3 \\
PhotoAssign           &  0.8 &  1.4 &   1.0 &   0.2 &   0.9 \\
PhotoProcessor        & 32.6 & 21.4 &  42.2 &  53.3 &  52.9 \\
PhotoReceive          &  6.5 &  0.1 &   3.1 &   4.6 &   2.6 \\
PhotoReport           &  2.0 &  6.5 &  11.1 &  16.2 &  18.4 \\
ProductCatalogApi     & 10.2 &  9.7 &   1.0 &   5.2 &   5.6 \\
ProductCatalogBuilder & 11.0 &  7.3 &   2.8 &  14.5 &  16.3 \\
\midrule
All functions                        &  9.8 &  6.9 &   9.4 &  14.5 &  14.8 \\
\bottomrule
\end{tabular}
\end{table}

\begin{figure*}[t!]
    \centering
    \begin{subfigure}[b]{0.33\textwidth}
        \centering
        \includegraphics[width=\textwidth]{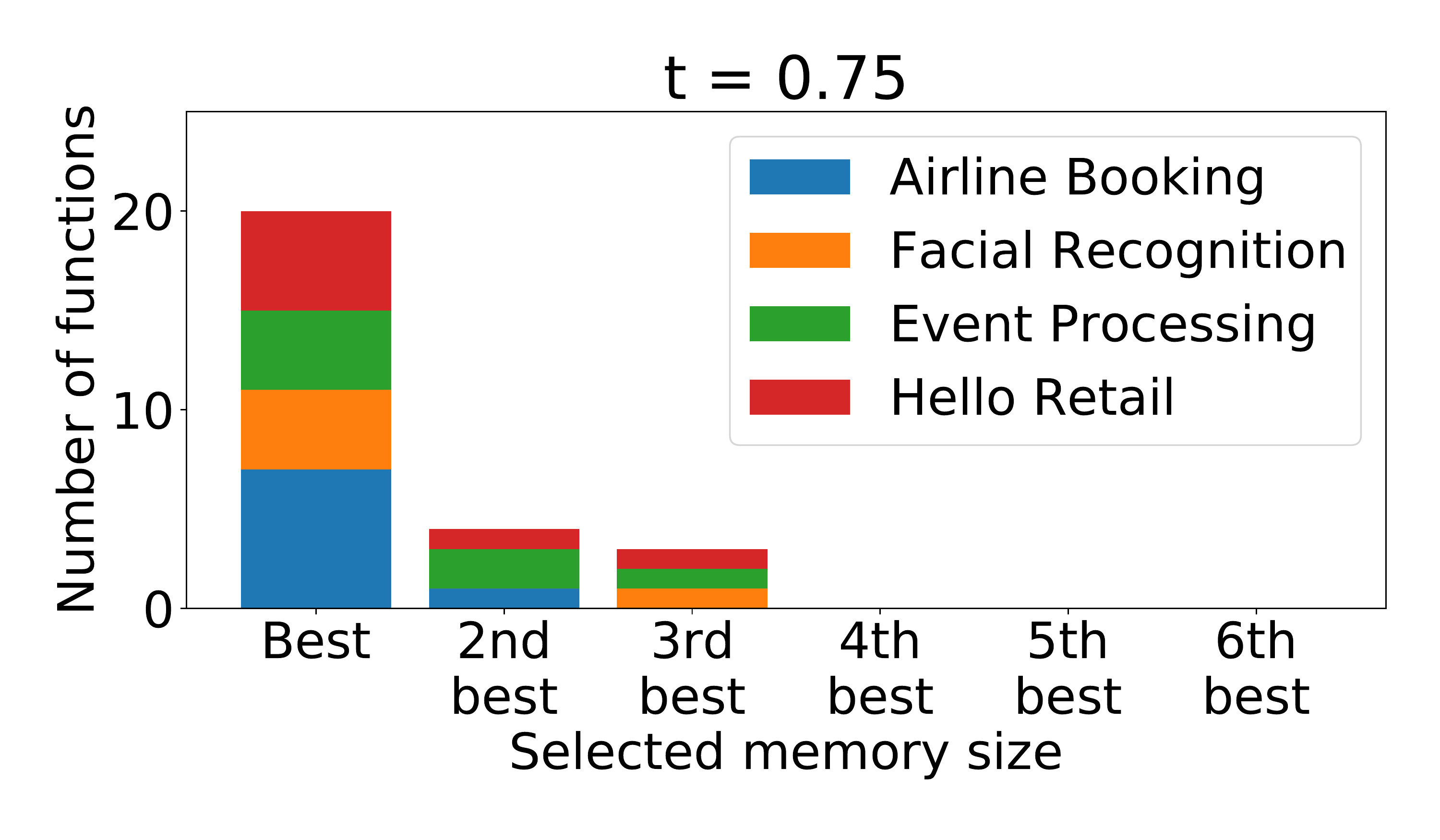}
    \end{subfigure}
    \begin{subfigure}[b]{0.33\textwidth}  
        \centering 
        \includegraphics[width=\textwidth]{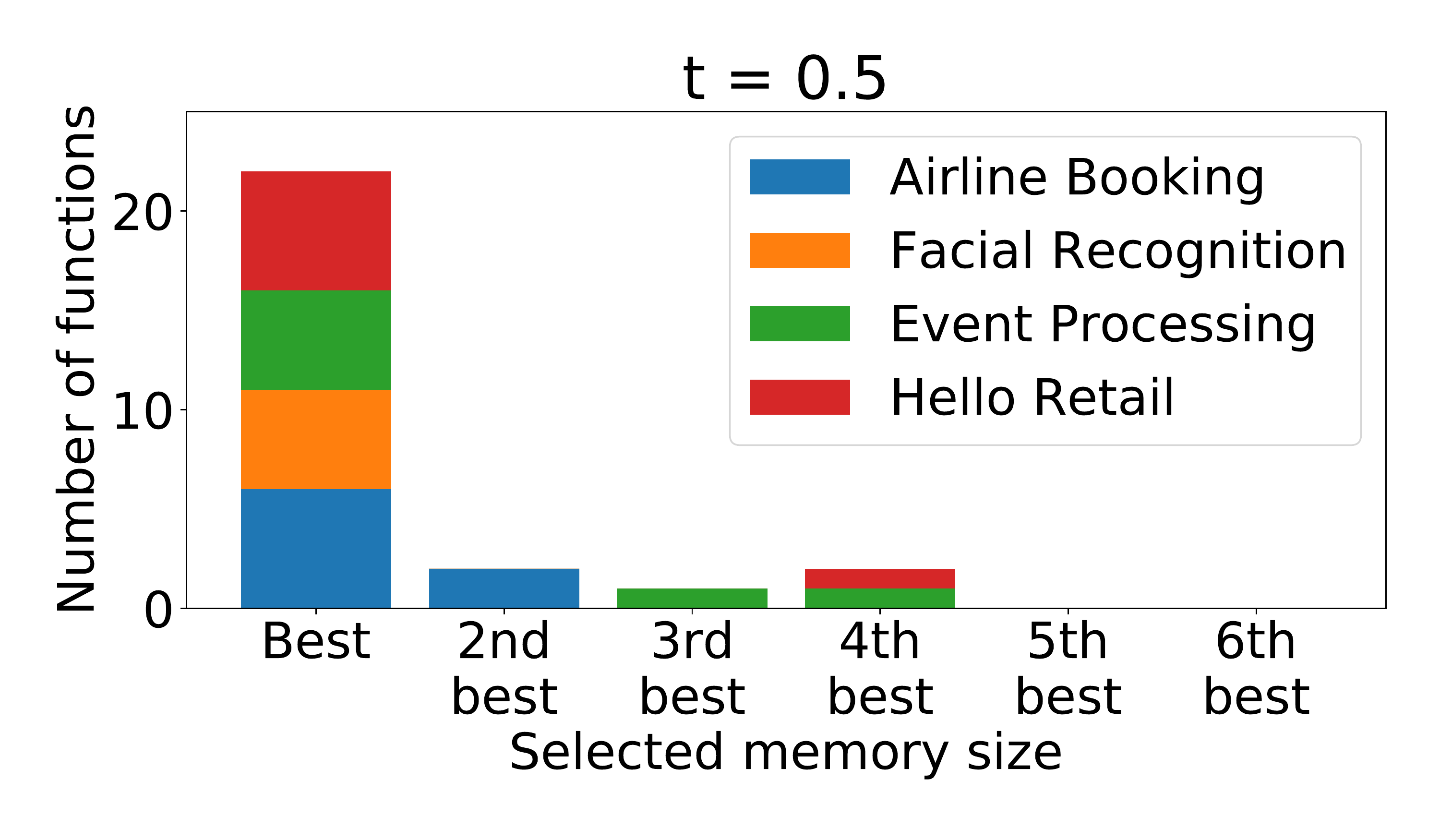}
    \end{subfigure}
    \begin{subfigure}[b]{0.33\textwidth}  
        \centering 
        \includegraphics[width=\textwidth]{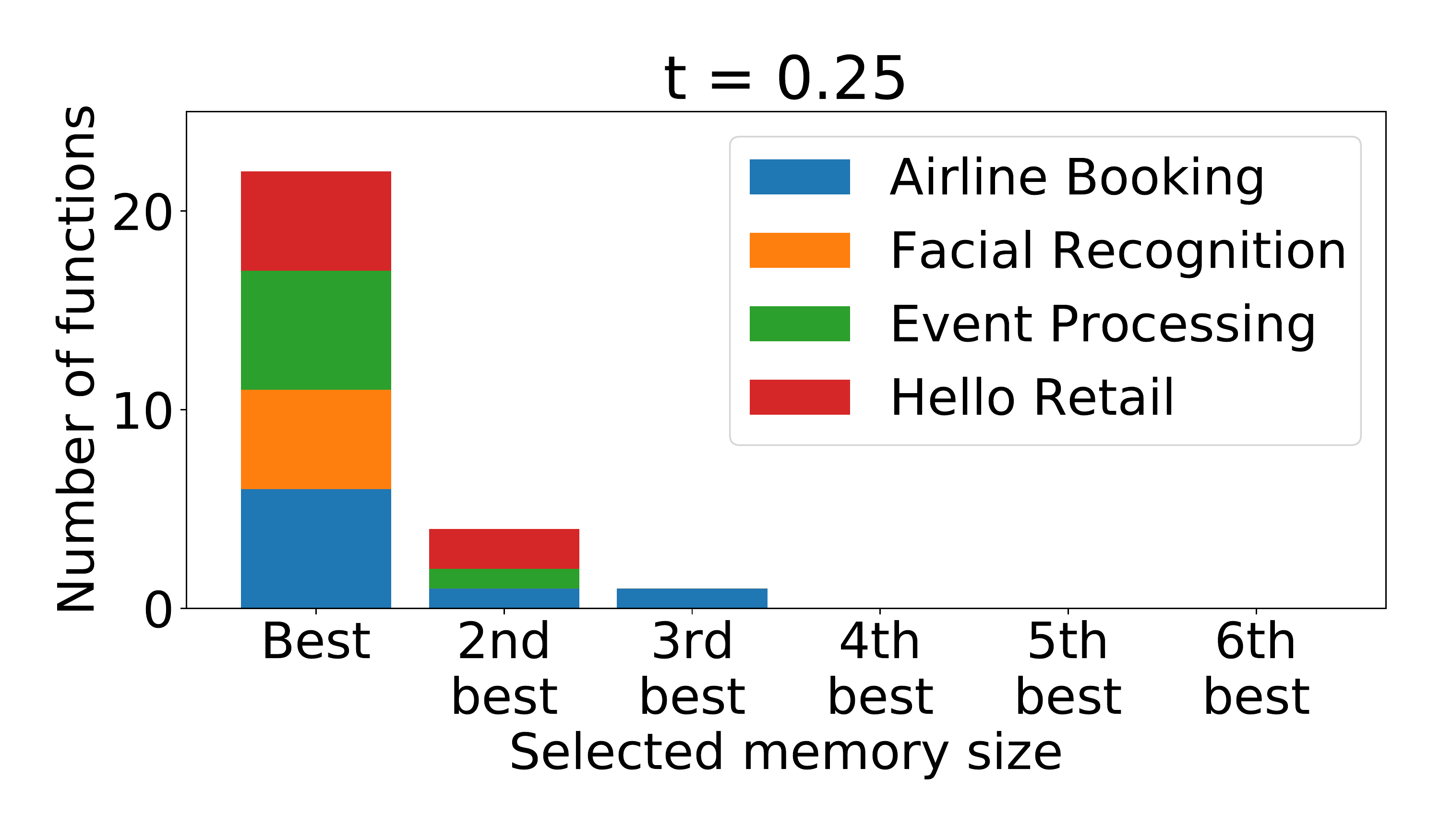}
    \end{subfigure}
    \caption{Number of functions for which our approach selects the X best approach for three different tradeoff parameters.} 
    \label{fig:selectedmem}
\end{figure*}
\begin{table*}[t!]
\small
\centering
\caption{Cost savings and application speedup after optimizing the four applications using our approach.}
\label{tab:benefit}
\begin{tabular}{lrr|rr|rr}
\toprule
\multirow{2}{*}{Application} & \multicolumn{2}{c}{t = 0.75} & \multicolumn{2}{c}{t = 0.5} & \multicolumn{2}{c}{t = 0.25} \\
&  Cost savings  &  Speedup & Cost savings  &  Speedup & Cost savings  &  Speedup\\
\midrule
Airline Booking   &  15.6\% &  28.5\% &     4.5\% &  31.9\%&        -12.3\% &  34.1\%\\
Facial Recognition  &  -2.9\% & 67.5\% &    -2.9\% &  67.5\%&       -17.4\% &  70.6\%\\
Event Processing     &  2.8\% &  31.2\% &   -17.0\% &  47.8\% &     -30.5\% &  57.5\%\\
Hello Retail     &  -8.4\% &  41.1\% &      -32.4\% &  47.7\% &     -63.8\% &  55.7\%\\
\midrule
All Applications      &  2.6\% &  39.7\% &  -12.0\% &  46.7\% &     -31.3\% & 52.5\%\\
\bottomrule
\end{tabular}
\end{table*}

Table~\ref{tab:airline} shows the relative prediction error of the airline application for a base size of 256MB, which we identified as the preferred basesize as described in Section~\ref{sec:modeling}. Our approach is able to predict the execution time for 128MB and 512MB accurately with below 10\% relative error, and the relative prediction error increases to around 15\% for the execution time with 1024MB, 2048MB, and 3008MB. The predictions for the functions of the facial recognition application (Table~\ref{tab:face}) show similar prediction accuracies with 512 MB and 3008 MB below 10\% relative error and the other three memory sizes with below 15\% error. For the functions of the event processing application, the predictions for 128MB again exhibits a relative prediction error of around 10\%, however, the relative prediction error for 512MB, 1024MB, 2048MB, and 3008MB is quite large with 20-35\%, as shown in Table~\ref{tab:event}. This is mostly due to a single function where the predictions are over 100\% off. 
While the very high relative prediction error is partly due to the low absolute values at higher memory sizes (prediction $\sim$40ms, real $\sim$20ms), the approach also underestimates how well the function scales with additional resources. 
Finally, the prediction error for hello retail application is below 10\% for 128MB, 512MB, and 1024MB and below 15\% for 2048 MB and 3008 MB. Overall, our approach achieved an average prediction error of 15.3\%.
\newline \newline

\vspace{-15px}\noindent\textbf{RQ2:} \textit{Are the execution time predictions provided by our approach sufficient to determine the optimal memory size of serverless functions?} The goal of our approach is to select the optimal memory size for a serverless function after monitoring it with a single base size. Therefore, the goal of this experiment is to verify, whether our approach is applicable to find the optimal memory size based on the execution time predictions for the previously unobserved memory sizes.

To investigate if the predictions are accurate enough to determine the optimal memory size, we apply the optimization approach from Section~\ref{sec:optimization} using the execution time predictions and compare the selected memory size to the optimal memory size determined based on the measured execution time. We run the optimization for three different tradeoff parameters, $t=0.75$ which prioritizes cost, $t=0.5$ which shows no preference, and $t=0.25$ which prioritizes performance. Figure~\ref{fig:selectedmem} shows the ranking of the selected memory sizes for each tradeoff parameter grouped by application. For a tradeoff parameter of 0.75, our approach selects the optimal memory size for 74.0\% of the functions, for a tradeoff parameter of 0.5 it selects the optimal memory size for 81.4\% of the functions, and for a tradeoff parameter of 0.25, it also selects the optimal memory size for 81.4\% of all functions. If our approach does not select the optimal memory size, it usually selects the second-best memory size and only rarely the third or fourth-best memory size. %

Overall, our approach selects the optimal memory size for 79.0\% of the serverless functions and the second-best memory size for 12.3\% of the serverless function. This suggests that our approach is reliable enough for the automated memory size optimization of serverless functions.
\newline \newline

\vspace{-15px}\noindent\textbf{RQ3:} \textit{How large are the benefits in terms of decreased cost and execution time of our proposed approach?}
We have shown that our approach is able to select either the optimal or a close to optimal memory size. 
Next, we are investigating what the actual benefits of using the memory sizes selected by our approach are, i.e., how much costs can be saved and how much the function execution can be sped up.

To quantify these benefits, we calculate the relative change in cost and execution time between the memory size selected by our approach for the tradeoff factors of 0.75, 0.5, and 0.25. Table~\ref{tab:benefit} shows the average percentage cost savings and execution time speedup obtained by switching to the memory sizes recommended by our approach. 
For $t = 0.5$, the cost increase by 12.0\%, but the average function execution is speed up by 46.7\%.  If the optimization is configured to favor speed over cost ($t = 0.25$), the execution is sped up further (52.5\%) and the cost increases by 31.3\%. If the optimization is configured to favor cost over speed ($t = 0.75$), the cost savings increase to 2.6\% and the achieved speedup decreases by about seven percent to 39.7\%. This shows that the tradeoff parameter correctly influences the behavior of the optimization. As cost savings from the memory size optimization are generally lower than execution time speedups, it seems that $t = 0.75$ achieves the most balanced optimization result. Therefore, we would recommend using this tradeoff factor for the automated memory size optimization.

To summarize, using our approach in this balanced configuration to optimize the memory size of four realistic serverless functions saves on average 2.6\% costs and speeds up the functions by 39.7\%. This highlights the importance of selecting an appropriate memory size and therefore the benefits of an approach for the automated memory size optimization.

\section{Limitations}
\label{Limitations}
While we consider our approach a significant improvement over the current state of the art, there still are limitations and threats to the validity to be discussed.

First, we limited the problem along two dimensions to reduce the cost associated with generating the synthetic dataset. AWS actually supports adjusting the memory size from 128MB to 3008MB in 64MB increments, whereas the dataset in this paper is limited to only six different memory sizes. However, the approach from~\cite{ali2020batch} could be used to interpolate the values for the 64MB increments. Second, our approach currently only supports a single cloud provider and a single programming language. We are confident that transferring the approach to other providers and languages does not pose any conceptual challenges and only requires an extension of the synthetic dataset as no Node.js-specific metrics were used.

Second, we did not evaluate the performance overhead caused by our resource consumption monitoring. While this overhead does not impact the measured metrics and execution times, it might hinder adoption in practice. However, there is already a commercial monitoring solution that tracks the CPU usage and network activity of serverless functions using Lambda Extensions, which indicates that the overhead of monitoring system-level metrics for serverless functions is reasonable~\cite{Yan2020Extensions}. Therefore, monitoring six system-level metrics for about ten minutes should not cause any issues.

Third, a shift in the workload of an application can change the performance properties of a serverless function, e.g., the workload becomes substantially burstier, which causes more cold starts or the payload size increases, which causes longer execution times. These workload shifts would also change the resource consumption metrics, so our model could be used to predict the optimal memory size for the changed function behavior again.

Finally, serverless platforms are introducing new features and performance improvements regularly, which raises the question of the longevity of our model. The measurements for the hello retail application were conducted nine months after the training dataset was created and there is no significant deterioration in prediction accuracy pointing. However, there might be breaking changes on a provider side that would invalidate our model. To avoid having to regenerate the full training dataset, one could explore transfer learning techniques that freeze the initial layers of our model and retrain only with a much smaller new dataset.

\section{Related Work}
\label{RelatedWork}
The existing work related to this paper can be grouped into three categories: measuring the impact of different memory sizes on the cost and execution time of serverless functions, general approaches for the cost optimization of serverless functions, and other approaches aiming to optimize the memory size of serverless functions.

\paragraph{Measuring the impact of different memory sizes}
\label{rel2}
Zhang et al. analyze the impact of different configuration options of serverless functions on their cost and performance~\cite{serverlessVideoProc}. They observe that the impact of memory sizes is non-trivial and highly
related to the workload. Further, they conclude that dynamic profiling is necessary to find the best memory configuration.
In 2018, Wang et al. conducted one of the largest measurement studies of serverless functions to date~\cite{PeekingBehindTheCurtains}. They find that memory size impacts not only the function execution time but also the cold start duration. Further, they quantify how I/O and network capacity increase with larger function sizes.
Figiela et al. introduce a benchmarking framework for serverless functions~\cite{figiela2018performance}. Using this framework, they analyze the execution time distribution of multiple functions for different memory sizes on AWS, GCloud, and IBM Cloud. Here, they find that the impact of memory size on execution time differs between providers and that finding the best performing memory size is non-trivial.
Back et al. evaluate the performance and cost model of public and private serverless function platforms using microbenchmarks~\cite{back2018using}. They find that the relation between memory size and cost is not linear and depends on the cloud provider.

For an in-depth overview of the current state of serverless function benchmarking, we refer to a recent survey by Scheuner et al.~\cite{scheuner2020function}.

\paragraph{Cost optimization of serverless functions}
Eismann et al.~\cite{predictingServerlessWorkflows} proposed an approach to optimize the costs of serverless workflows. This approach uses mixture density models to predict the execution time distribution of a function based on its input parameters. One of the limitations of this approach is that it assumes a fixed memory size. 
In the work of Boza et al.~\cite{serverlessSimBudget}, an approach using model-based simulations to compare the costs of reserved VMs, on-demand VMs, and serverless functions is introduced. The authors propose to model serverless functions as $M(t)/M/\infty$ queues, where the users need to provide the used memory size and resulting average service time as configuration parameters.
Elgmal et al.~\cite{costless} propose an approach to optimize the costs of serverless workflows by deciding whether to fuse multiple functions into a larger function and which memory limit should be allocated to a serverless function. This approach also requires users to provide the average execution time for each memory size as input.
Gunasekaran et al.~\cite{spock} propose to use serverless functions in combination with VM-based hosting to enable SLO and cost-aware resource procurement. To enable the cost-aware decision making between VM-based hosting and serverless functions, this approach also requires the user to provide the average execution time for the selected memory size as input.

To summarize, these existing cost optimization approaches either assume a fixed memory size or require the mean execution time for different memory sizes as user input. The approach presented in this paper can augment these existing cost optimization approaches by providing information about the average execution time for different memory sizes.

\paragraph{Optimizing the memory size of serverless functions}
The first approach is the AWS power tuning tool, a popular open-source tool that measures the impact of different memory sizes on the execution time and cost of a serverless function~\cite{Casalboni2020Power}. A step function workflow coordinates the deployment, performance measurement, and result collection for a set of predefined memory configurations.
The second approach is COSE, which aims to reduce the number of required measurements using Bayesian optimization~\cite{Akhtar2020COSE}. By learning a performance model describing the relationship between memory size and execution time based on fewer measurement points, COSE can reduce the required performance measurements.
Lastly, BATCH is a framework for efficient machine learning serving on serverless platforms~\cite{ali2020batch}. It relies on a profiler that measures a subset of the potential configurations (i.e., memory size, batch size, and timeout) and employs a multivariable polynomial regression model to estimate the performance of the remaining configurations.

To summarize, all existing approaches for the memory size optimization of serverless functions, combine sparse measurements with interpolation/modeling to determine the optimal memory size, an approach that is commonly applied for data analytics systems~\cite{Venkataraman2016, Alipourfard2017, Herodotou2011}.
However, all existing approaches require measuring multiple sizes. 

\section{Replication Package}
\label{ReplicationPackage}
To facilitate replication of our results and extensions of our approach, we attempt to provide an extensive three-part replication package. This is in line with a recent guideline on reproducible measurements in cloud environments~\cite{Papadopoulos2019methodological}.

The first part of our replication package consists of docker containers containing the benchmark harnesses for the airline booking, facial recognition, and event processing applications that require only the injection of valid AWS credentials to reproduce the measurement results shown in our evaluation. 
The second part of our replication package consists of the function generator and resource consumption monitoring introduced in this paper alongside instructions to reproduce the measurements to create our training dataset. 
The third part is a CodeOcean capsule containing the generated dataset covering 12\,000 performance measurements (216\,000\,000 lambda executions) as well as the implementation of the multi-target regression modeling used in our approach. This CodeOcean capsule enables a 1-click replication of all analyses conducted in the scope of this work, and all tables/figures presented in the evaluation. The replication package is available in form of a Github repository\footnote{\url{https://github.com/Sizeless/ReplicationPackage}}.

\section{Conclusion}
\label{Conclusion}
Serverless functions automate resource provisioning, deployment, instance management, and auto-scaling. The last resource management task that developers are still in charge of is resource sizing, i.e., selecting how much resources are allocated to each worker instance.  

In this paper, we introduced an approach to predict the optimal resource size of serverless functions using monitoring data of a single memory size. First, we introduced a synthetic function generator and a resource consumption monitoring approach. Using these, we generated a large dataset on how functions with different resource consumption behavior scale with increasing memory sizes. Based on this dataset, we trained a multi-target regression model capable of predicting the execution time of a serverless function for all memory sizes based on monitoring data for a single memory size. These predictions then enable the automated optimization of a serverless function's memory size.

This is the first approach that requires only passive monitoring data from an already deployed function, instead of active performance measurements. By automating the memory size optimization for serverless functions, our approach removes the last resource management task that developers still need to deal with in serverless functions and thus makes serverless functions truly serverless. 
In our evaluation, our approach predicts the execution time of the other memory sizes with an average prediction error of 15.3\% based on monitoring data for a single memory size. It selected the optimal memory size for 79.0\% of the serverless functions and the second-best memory size for 12.3\% of the serverless functions, which results in an average speedup of 39.7\% while simultaneously decreasing average costs by 2.6\%. 
\section*{Acknowledgments}

This work was supported by the AWS Cloud Credits for Research program. The authors would like to thank Alex Casalboni for providing the measurement data behind the motivating examples.

\bibliographystyle{ACM-Reference-Format}
\bibliography{bibliography}

\end{document}